\documentclass{article}
\usepackage{amsmath}
\usepackage{amsfonts}
\usepackage{amssymb}
\usepackage{graphicx}
\newcommand{\bsigma}{\mbox{\boldmath $\sigma$}}
\newcommand{\btau}{\mbox{\boldmath $\tau$}}
\newcommand{\car}{$^{12}$C\,\,}
\newcommand{\oxy}{$^{16}$O\,\,}

\newcommand{\anu}{\overline \nu}
\newcommand{\beq} {\begin{equation}}
\newcommand{\eeq} {\end{equation}}

\newcommand{\bqu}{{\bf q}}

\begin{document}
\begin{center}
{\LARGE \bf Excitation of nuclear giant resonances 
 in neutrino scattering off nuclei} 
\vskip 0.5 cm
{\Large Antonio Botrugno and Giampaolo Co'}
\vskip 0.5 cm
{\Large Dipartimento di Fisica, Universit\`a di Lecce \\
and \\
Istituto Nazionale di Fisica Nucleare  sez. di Lecce, \\
I-73100 Lecce, Italy}

\end{center}
\begin{abstract}
  The cross sections for neutrino scattering off the \car and \oxy
  nuclei are calculated within the framework of the continuum Random
  Phase Approximation.  A model to consider also the final state
  interactions is developed.  Total charge-conserving and
  charge-exchange cross sections for both electron neutrinos and
  antineutrinos have been calculated up to projectile energies of 100
  MeV.  The sensitivity of the cross sections to the residual
  interaction and to the final state interactions is investigated.  A
  direct comparison between neutrino and electron scattering cross
  sections calculated under the same kinematic conditions is
  presented. We found remarkable differences between electromagnetic
  and weak nuclear responses.  The model is applied to describe cross
  sections of neutrinos produced by muon decay at rest and in
  supernovae explosions.
\end{abstract} 
\section{Introduction}
There are various reasons to study the interactions of neutrinos with
complex nuclei.  A first set of motivations is related to the use of
the nucleus as a detector to investigate the neutrino properties
\cite{nuint}, for example, the parameters of the flavor oscillations.
A second group of motivations is related to use of the neutrinos as a
probe in astrophysics \cite{cav04}, and, eventually, in geophysics
\cite{fio03}. Here the importance of the interaction with nuclei is
not limited to the detection of the neutrinos, but it is also related
to the study of the processes generating them, and to their scattering
inside matter (stars, supernovae or earth). Another set of
motivations is related to the use of neutrinos to probe the structure
of the nucleons and of the nuclei.  There is great interest for the
possibility of studying the strangeness content of the nucleon with
neutrino scattering \cite{gar92,mus94,alb02}.  Furthermore, neutrinos
excite nuclear modes not accessible to the electromagnetic probes, and
this allows the study of the characteristics of the nuclear dynamics, and
of the nuclear interaction, usually hidden in other processes.

The phenomena above mentioned, involve neutrinos energies varying from
a few MeV up to thousand of TeV.  In this paper we study processes with
neutrinos energies limited to a few tens of MeV.  In this case the
nucleon degrees of freedom are not excited and we shall be concerned
only about the excitation of the nucleus.

Neutrinos with few tens of MeV can excite the nucleus above the
nucleon emission threshold, in the region of the giant resonances.
The Continuum Random Phase Approximation (CRPA) is the theory most
often used to describe the excitation of the nuclear giant resonances.
The calculations discussed here have been done by solving the CRPA
equations with the Fourier-Bessel technique of Ref. \cite{deh82}.  The
CRPA is able to describe, and predict, the position of the various
resonances. In the comparison with the experimental data the CRPA
cross sections are too large and their widths too narrow. There are
good indications that these problems can be solved by extending the
RPA configuration space which considers only elementary excitations of
one-particle one-hole (1p-1h) type \cite{dro90,kam04}. We call Final
State Interactions (FSI) the effects beyond CRPA, to allow for further
re-interaction of the emitted nucleon with the remaining nucleus.  We
have improved the CRPA description of the total photo-absorption cross
sections \cite{ahr75} by developing a phenomenological model that
considers the FSI.  In this article we present the results obtained by
applying the CRPA plus the FSI model to the calculations of
neutrino-nucleus cross sections.  The primary aim of this work is the
study of the consequences of the uncertainties of the nuclear
structure model on the description of the neutrinos cross sections.

Even though the formalism we have developed is quite general, we have
limited our investigation to the study of the interaction of electron
neutrinos, and antineutrinos, with the \car and \oxy nuclei.  The two
nuclei we have selected are doubly magic nuclei where the CRPA theory
is more successful.  Furthermore, they are the most important nuclei in
liquid scintillators (\car) or water (\oxy) detectors.

In Sect. \ref{ssec:crpa} we discuss in some detail the choice of the
CRPA input parameters.  The model developed to treat the FSI is
presented in Sect. \ref{ssec:fold}.  Our results are presented in
Sect. \ref{sec:res}.  We first discuss some general features of the
neutrino-nucleus cross sections, then we compare electron ad neutrino
cross sections calculated with the same kinematics.  The total cross
sections on the \car and \oxy nuclei are shown in Sect.
\ref{ssec:totcs}.  In Sect. \ref{ssec:appl} we present the results
obtained by applying our model to the case of neutrinos produced by
muon decay at rest, and in supernova explosion.

\section{The model}
\label{sec:nuclmod}
In this work we consider the inelastic neutrino-nucleus scattering
processes involving both charged current (CC) and neutral current (NC)
reactions. The neutrino-nucleus cross section is obtained in plane
wave Born approximation and for point-like interaction \cite{wal75}. In
the case of charged leptons we consider the Coulomb distortion by using
a Fermi function \cite{bla52,eng98}. 
The expression of the cross section has been obtained by using the
standard trace techniques. We made a non relativistic reduction of the
weak current operators, and a multipole expansion of the transition
amplitudes \cite{wal75}. For the nucleon axial form factor we used a
dipole expression with axial mass $M_A$= 1014 MeV \cite{ber02}.  

This derivation of the cross section is a rather standard one
\cite{wal75}, and we do not present it here.  A detailed description
can be found in Ref.  \cite{bot04}.  It may be worth, however, to
discuss the final expression of the neutrino-nucleus inclusive cross
section, by pointing out the differences with the electron scattering
cross section. Under the approximations above mentioned, the the
electron scattering cross section can be written as a sum of a
longitudinal response, i.e. parallel to the momentum transfer
direction, produced by the charge operator, and a transverse response,
related to the current operator.  Also the neutrino-nucleus cross
section can be separated into a longitudinal and transverse responses.
The novelties in the neutrino scattering are originated by the
presence of an axial-vector term of the weak interaction operators in
addition to the vector term which is analogous to that of the
electromagnetic four-current. For this reason, in the neutrino
scattering case, the transverse response is composed by a purely
vector term, an axial-vector one and the interference between them.
The longitudinal response is composed by a vector Coulomb-like term
and two axial terms which interfere with each other.  The presence of
a single vector term in the longitudinal response is due to the fact
that we used the conserved vector current hypothesis.  Always in the
longitudinal response, the vector and the axial terms do not interfere
since they excite states of different parity. Specifically, the vector
current excites natural parity states only, while the axial current
only unnatural parity states \cite{wal75}.

\subsection{Continuum Random Phase Approximation}
\label{ssec:crpa}
The evaluation of the hadronic transition matrix elements,
requires the knowledge of the wave functions
describing the ground state and the excited states of the nucleus. As
already anticipated in the introduction, we work in the framework of
the CRPA theory.  The secular equations of the theory are solved
by using the Fourier-Bessel expansion technique developed in
\cite{deh82} and used in the description of various electron
scattering processes \cite{co84,co85,co88}.  We have extended the
basic formalism of Ref.  \cite{deh82} to treat charge-exchange
excitations.  This extension is straightforward,
therefore, we do not present here a detailed description of the theory
which can be found elsewhere \cite{bot04}.  We rather prefer to make a
discussion of the input parameters necessary to solve the CRPA
equations.

The first input required by the CRPA is the mean field used to
generate the configuration space of single particle states.  The
Fourier-Bessel formalism requires single particle states both in the
discrete and in the continuum part of the spectrum. For both cases we
use the same Woods-Saxon potential of the form:
\begin{equation}
U\left( r\right) =U_{C}(r)+U_{SO}(r)+U_{Coul}\left( r\right)
\label{eq:ws1}
\end{equation}
with
\begin{equation}
U_{C}(r)=\frac{V}{1+exp[(r-R)/a]} 
\label{eq:ws2}
\end{equation}
\begin{equation}
U_{SO}(r)=\left( \frac{\hbar }{m_{\pi }c}\right)^{2}
\frac{1}{r}\frac{d}{dr}
\left( \frac{V_{s}}{1+exp[(r-R_s)/a_s]}\right) \mathbf{l\cdot s} 
\label{eq:ws3}
\end{equation}
where $m_{\pi}$ is the pion mass, and $U_{Coul}$ is the Coulomb
potential, active only for protons, generated by a uniformly charged
sphere of radius $R_c$ of total charge $Z-1$.

The parameters used for the \car and \oxy nuclei are shown in Table
\ref{tab:woodpar}.  They have been chosen to reproduce at best the
single particle energies around the Fermi surface and empirical charge
density distributions \cite{dej87}.  The \car para\-mete\-riza\-tion is
different from that of Ref. \cite{co84}, more commonly used in the
literature. With this new set of parameters the \car empirical charge
distribution is much better reproduced.

The single particle energies are a crucial ingredient for the solution
of the CRPA equations. In the spirit of the Landau-Migdal theory of
the finite Fermi systems \cite{mig67}, we used, when available, the
experimental single particle energies \cite{led78}. 
In Table \ref{tab:spe} we show
the single particle states forming the discrete basis of our
configuration space, together with the theoretical and experimental
energies.  In this table, the proton 1d5/2 and the neutron 1d3/2
states are included in \car.  In our calculations, these states are
in the continuum but the widths of their resonances are so narrow that,
for computational reasons, it is more convenient to consider them as
bound states with positive energy.  The full space used in our
calculations is obtained by adding all the continuum states allowing
the angular momentum couplings with the hole states up to total
angular momentum $J$=16.

The other ingredient required by the CRPA calculations is the residual
interaction. We studied the sensitivity of our results to the choice
of the residual particle-hole ($p-h$) interaction by making CRPA
calculations with three different forces. Their general expression, in
momentum space, is:
\begin{eqnarray}
\nonumber
V^{ph} (k,\rho) &=& \mathcal {K}
 \left[ \right.
  F(k,\rho) 
 + F'(k,\rho) \, \btau(1) \cdot \btau(2) 
\\
&+&
 G(k) \, \bsigma(1) \cdot \bsigma(2)
 + G'(k) \, \bsigma(1) \cdot \bsigma(2) \, \btau(1) \cdot
 \btau(2)
\left. \right]
\label{eq:force}
\end{eqnarray}
where $k$ is the relative momentum of the $p-h$ pair, and
$\mathcal {K}$ a normalization constant. 
The first two terms have a linear dependence from the nuclear density
$\rho$ \cite{rin78}:
\begin{equation}
 F(k,\rho) =  F^{ext}(k,\rho)
+\left[ F^{int}(k,\rho) -  F^{ext}(k,\rho) 
\right] \frac { \rho(r) }{\rho(0)}
\label{eq:frho}
\end{equation}
In our calculations we used 
\begin{equation}
\rho ( r ) =\frac {1} { 1 + exp[(r-R)/a] }
\end{equation}
with $a$ =  0.5  and 0.6 fm and $R$ = 3.0 and 2.71 fm for 
\car and \oxy, respectively.

In our calculations we used two interactions of contact type, or
zero-range interactions.  This means that the functions
$F$, $F'$, $G$,and $G'$ of Eqs. (\ref{eq:force}) and (\ref{eq:frho}) are
constant in $k$.  The values of these constants are given in Table
\ref{tab:force}, and they are labeled LM1 and LM2.  The parameters of
the LM1 force have been fixed in \cite{rin78} to reproduce the
muonic-atom nuclear polarization energy shifts in the region of the
$^{208}$Pb nucleus. This parameterization of the interaction has been
used in the literature to describe electromagnetic excitations of
doubly magic nuclei \cite{deh82,co84,co85}. The parameters of the
second zero-range interaction, LM2, have been fixed in \cite{bak97} to
describe the nuclear spin responses of \car.

\begin{center}
\begin{figure}
\includegraphics [bb= 20 30 600 800,angle=0,scale=0.70] 
                     {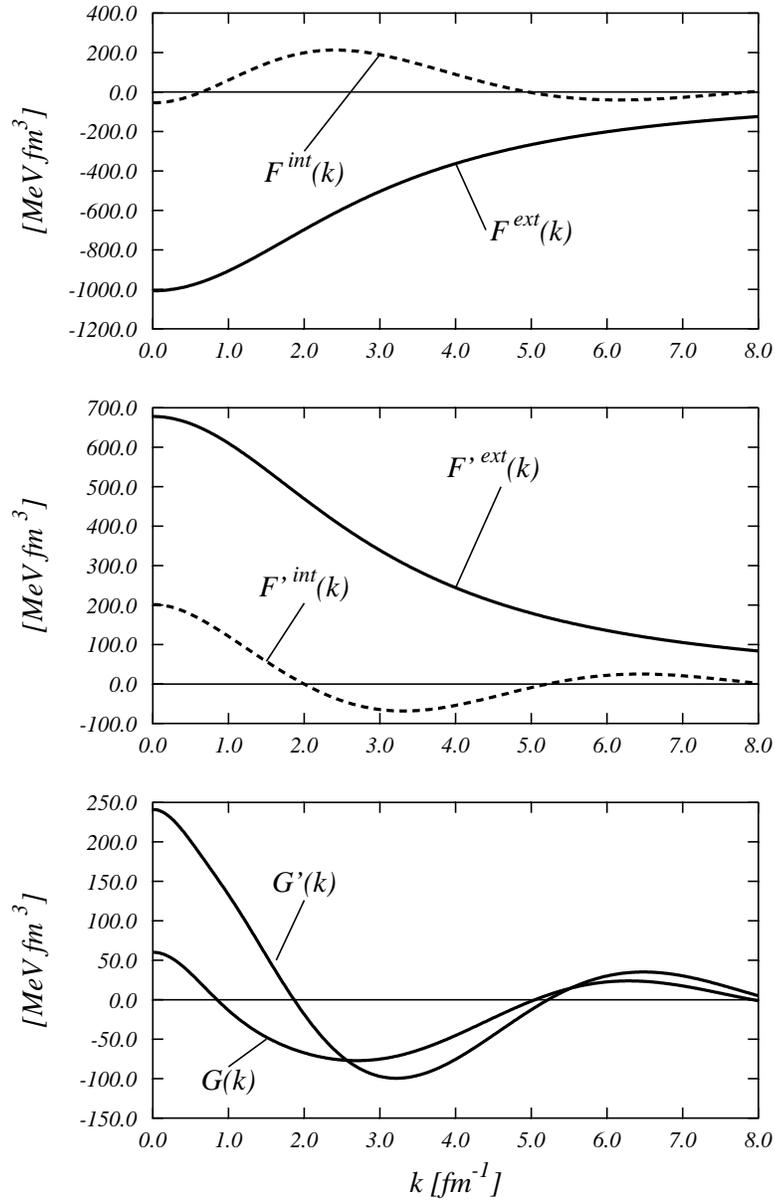} 
\vskip 0.1 cm
\caption{\small The various terms of the PP interaction as a function
                of the nucleons pair relative momentum $k$.
}
\label{fig:PP}
\end{figure}
\end{center}

In addition to these two forces we used the polarization potential
(PP) of Ref. \cite{pin88}. This is an effective finite-range
interaction containing, in an average way, the contribution of the
exchange diagrams.  The parameters of this force have been fixed
to reproduce some static properties of nuclear matter. The validity of
this parameterization has been tested against the results of
microscopic calculations of nuclear matter density responses
\cite{fan87}.  The PP interaction has been used in the description of
the electron scattering quasi-elastic responses \cite{co88,ama01}.  The
values of the Landau parameters, obtained in the limit $k \rightarrow
0$, are given in Table \ref{tab:force}. The behavior of the various
terms of the force as a function of the relative momentum of the $p-h$
pair $k$ is given in Fig. \ref{fig:PP}.

The values of $F$, $F'$, $G$,and $G'$ presented in Table
\ref{tab:force} and in Fig. \ref{fig:PP}, have been fixed in specific
contexts.  Since we want to use these interactions in a different
context, with a very pragmatical approach, in the spirit of the
effective theories, we renormalize each interaction by multiplying it
with a constant $\mathcal K$, whose values are given in Table
\ref{tab:force}.  In the case of \oxy the values of the constant have
been fixed to reproduce the energy of the collective low-lying 3$^-$
state at 6.13 MeV. For the \car nucleus the constants have been fixed
to reproduce the position of the main peak of the total photo
absorption cross section at 22.5 MeV. Only the LM2 force in \car has
not been renormalized, since in \cite{bak97} it was specifically tuned
to treat this nucleus, even though the RPA calculations where
performed in a discrete single particle basis.

The performances of our CRPA calculations have been tested by
comparing the total photo-absorption cross sections with the data
measured in Mainz \cite{ahr75}. This comparison is shown in the panels
(a) and (c) of Fig. \ref{fig:sr}.  In our CRPA calculations the
multipole excitations up to 3$^-$ have been considered, even though
the 1$^-$ excitation contributes for more than the 97\% of the cross
sections shown in the figure.

In \oxy the LM1 and PP cross sections almost overlap, while the LM2 is
slightly more attractive. In any case the position of the peak is
quite well reproduced.  As already mentioned, in the \car case, the
good description of the main peak position provided by the LM1 and PP
cross sections has been obtained by changing the normalization
constant $\mathcal K$.  The differences between the two forces becomes
evident when observing the other structures of the cross sections.
The sharp peak at 17.8 MeV due to the neutron (1p3/2 $^{-1}$, 2s1/2)
transition is more strongly excited by the LM1 force.  The large peak
above 30 MeV, produced by the opening of the 1s1/2 emission channel,
appears at lower energy for the PP force. The LM2 force is more
attractive than the other two, also in the \car case.

The interactions we have adopted, especially LM1 and PP, have been
tuned to describe natural parity excitations. These are the most
relevant excitations induced by electromagnetic probes. On the other
hand, neutrinos are not so selective, and excite indifferently natural
and unnatural parity states. This point will be discussed with more
details in Sect. \ref{ssec:evsnu}. 

We have investigated the reliability of our approach in the
description of the unnatural parity excitations.  As example of the
results of our tests, we show in Table \ref{tab:oneplus} the excitation
energies of the 1$^+$ excitations in \car charge-conserving and
charge-exchange transitions.  The agreement with the experimental
energies \cite{led78} is not very satisfactory, even though the
general behavior is respected.  It is remarkable that the three
forces produce very similar results for the charge-exchange
transitions, contrary to what happens in the charge-conserving
ones. It is well known in the literature that the 
description of the unnatural parity excitations improves when 
tensor terms are included in the $p-h$ interaction \cite{spe80}.

\subsection{The FSI model}
\label{ssec:fold}

\begin{center}
\begin{figure}
\includegraphics [angle=90, scale=0.5] 
{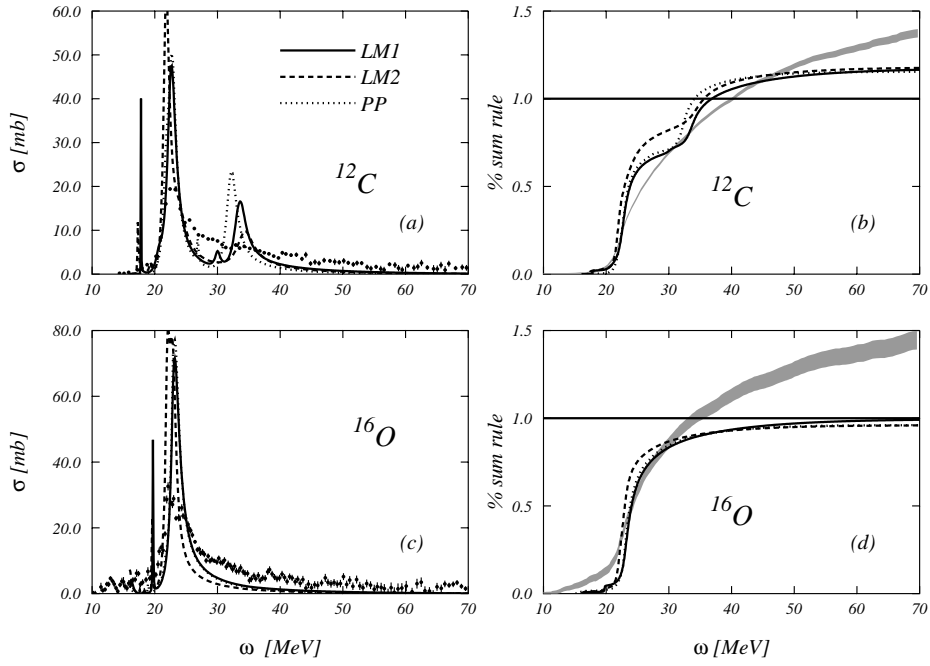}
\vskip 0.1 cm
\caption{\small Panels (a) and (c): total 
  photo-absorption cross sections of
  \car and \oxy calculated with the CRPA, compared to the experimental
  data of Ref. \protect\cite{ahr75}.  In the
  panels (b) and (d) we show the sum rules
  calculated as integral of the cross sections shown in panels (a) and
  (c), see Eq. (\ref{eq:sr}),
   and normalized to the classical value of the Thomas-Reiche-Kuhn
  sum rule. The gray bands indicate the experimental sum rules. The LM1,
  LM2, and PP labels refer to the three different residual
  interactions used in our work.
}
\label{fig:sr}
\end{figure}
\end{center}

The limits of the CRPA theory in describing the excitation of the
nuclear giant resonances, are evident in Fig.  \ref{fig:sr}. While the
positions of the resonances are rather well reproduced, the cross
sections are overestimated, and their widths underestimated.  The use
in the calculations of different interactions does not substantially
modify the behavior of the CRPA results.  This indicates that the above
mentioned problems are not solvable by changing the input of the
calculations, but they are related to some of the intrinsic
approximations connected to the RPA theory. In the panels (b) and (d)
of Fig. \ref{fig:sr}, we show the behavior of the sum rule integral
\begin{equation}
SR(\omega) = \int_0^\omega d E \,\, \sigma(E) 
\label{eq:sr}
\end{equation}
as a function of $\omega$. The various curves in the panels (b) and
(d) of the figure have been normalized to the classical value of the
Thomas-Reiche-Kuhn (TRK) sum rule: $ 60 N Z / A$ mb. The sum rule
integrals of the experimental cross sections are shown by the gray
bands.

While all the CRPA results in \oxy are below the TRK value, the \car
asymptotic values are above this limit. The presence of the $F'$ and
$G'$ functions in our $p-h$ interactions violate the hamiltonian isospin
independence hypothesis used to obtain the TRK sum rule
values. The effects of these terms of the interaction are larger in
\car than in \oxy. A more detailed investigation of this point is
beyond the scope of the present article. The information we want to
give in this context is that all the CRPA calculations give asymptotic
values of the sum rules smaller than those obtained by integrating the
experimental cross sections.  There is a certain agreement between
CRPA results and empirical values up to about 40 MeV in \car and 30
MeV in \oxy but for larger energies, the experimental sum rules are
much larger than both the TRK values, and the CRPA curves. There are
effects beyond those described by the CRPA, probably short-range
correlations effects \cite{fab85}.

As we have already stated in the introduction, the main limitation of
the CRPA theory seems to be the restriction of a configuration space
where only 1p-1h excitation pairs are considered. The extension of
this configuration space, to include more complicated excitations, for
example 2p-2h pairs, seems to improve the agreement with the
data \cite{dro90,kam04}.  In order to take into account effects beyond
the RPA we extend a phenomenological model which
has been used to describe hadronic processes \cite{bak97,smi88} and
electron scattering cross sections in the quasi-elastic region
\cite{co88,ama01,ama94}.  The model assumes that excitations
beyond those considered by the RPA, act as intermediate doorway
states. Asymptotically, the nuclear excited states have one particle,
in the continuum, and one hole in the residual nucleus. A further
assumption consists in considering that, at sufficiently high
excitation energies, the contribution of the high-order excitations is
independent of the angular momentum of the nuclear excited state.
In the quasi-elastic region, the density of states is so high that a
specific nuclear excitation characterized by its angular momentum and
parity is not relevant any more.

Starting from these hypotheses, it is shown in Refs.
\cite{co88,smi88}, that the effects of these high-order excitations,
that we shall call henceforth FSI, can
be accounted for by a folding integral
\begin{equation}
S^{FSI}(q , \omega)  = \int_0^\infty dE \,\,\,
S^0(q, E) \left[ h(E,\omega)+h(E,-\omega) \right]
\label{eq:foldres}
\end{equation}
where the folding function is:
\begin{equation}
h(E,\omega)= \frac{1}{2\pi}
\frac{\Gamma(\omega)}
{ 
[ E-\omega-\Delta(\omega) ]^2 +
[ \Gamma(\omega)/2]^2 
}
\label{eq:folding}
\end{equation}
In the above equations $S^0$ indicates a known nuclear response for an
excitation energy $E$ and a momentum transfer $\bqu$. In our case
$S^0$ has been calculated within the CRPA theory. 

Fundamental ingredients of the folding 
model are the functions $\Delta(\omega)$ and $\Gamma (\omega)$ 
which are linked by a dispersion relation:
\begin{equation}
\Delta (\omega) = \frac{1}{2 \pi} P \int_{- \infty}^{+ \infty}
 d \omega ' \frac{\Gamma(\omega ')}{\omega ' - \omega} \,\, ,
\end{equation}
where we have indicated with $P$ the principal value integral. 
The problem is then limited to
the determination of the $\Gamma( \omega)$
function.  In a microscopic approach, this function has to be 
evaluated by
calculating all the many-particle many-hole transitions induced by the
residual interaction.  In practice, we extract $\Gamma( \omega)$ from
an empirical description of the imaginary part of the
single particle
self-energy.
The data we have considered for positive values of $\omega$ are those
related to the imaginary part of the optical potential whose
parameters have been fixed to fit nucleon--nucleus elastic scattering
cross sections. For negative values of $\omega$ we have considered the
energy width of the single particle levels, measured in knock--out
reactions like $(e,e'p)$ or $(p,d)$.  We have obtained $\Gamma
(\omega)$ by making the average of the single particle energy width
$\gamma(\omega)$ :
\begin{equation}
\Gamma (\omega) = \frac{1}{\omega} \int^\omega_0 d \varepsilon 
\left[  \gamma(\varepsilon+\omega) + \gamma(\varepsilon - \omega) \right] \,\, .
\label{eq:bgamma}
\end{equation}

The empirical values of the single particle widths \cite{mah82} have
been fitted by using the following expression of $\gamma$:
\begin{equation}
 \gamma ( \varepsilon) = A  
  \left( \frac{\varepsilon^2}{\varepsilon^2 + B^2} \right)
  \left( \frac{C^2}{\varepsilon^2+C^2} \right)
\label{eq:sgamma}
\end{equation}

Another correction to our CRPA calculations is related to
the non-locality of the mean field.
A simple way to take  into account this important correction
is the introduction of an effective mass of the nucleon $m^*$
\cite{mah82}. For a given value of $m^*/m$ the following scaling
relation holds:
\begin{equation}
  S_{m^*} (q, \omega) = \frac{m^*}{m} 
  S_{m} \left(q , \frac{m^*}{m}\, \omega \right) \,\,.
\label{eq:emass}
\end{equation}
Our FSI model conserves the sum rules. No further strength is added to
that provided by the RPA. The effect of the FSI is a redistribution of
the strength.

The parameterization of the $\gamma$ function and of $m^*$ given in
the lowest row of Table \ref{tab:gamma} allows us to reproduce the
electromagnetic quasi-elastic responses in various nuclei
\cite{co88,ama01,ama94,co02}. In the giant resonance region some of
the assumptions used to derive the model are not longer valid, for
example, the independence of the $\gamma$ function from the angular
momentum and the parity of nuclear excitation. We forced our model to
work also in the giant resonance region by considering an energy
dependence of the $A$, $B$ and $m^*$ parameters.  For energies greater
than 40 MeV, the values of these parameters are those used in the
quasi-elastic region. We fix the $A$, $B$ and $m^*$ values at 10 MeV,
and we let them evolve linearly up to 40 MeV, where they reach their
asymptotic values.  The values at $\varepsilon$ = 10 MeV have been
fixed to reproduce at best, within this model, the photo-absorption
data of Mainz \cite{ahr75}.  The full set of parameters is given in
Table  \ref{tab:gamma}.

\begin{center}
\begin{figure}
\includegraphics [angle=0, scale=0.7] 
{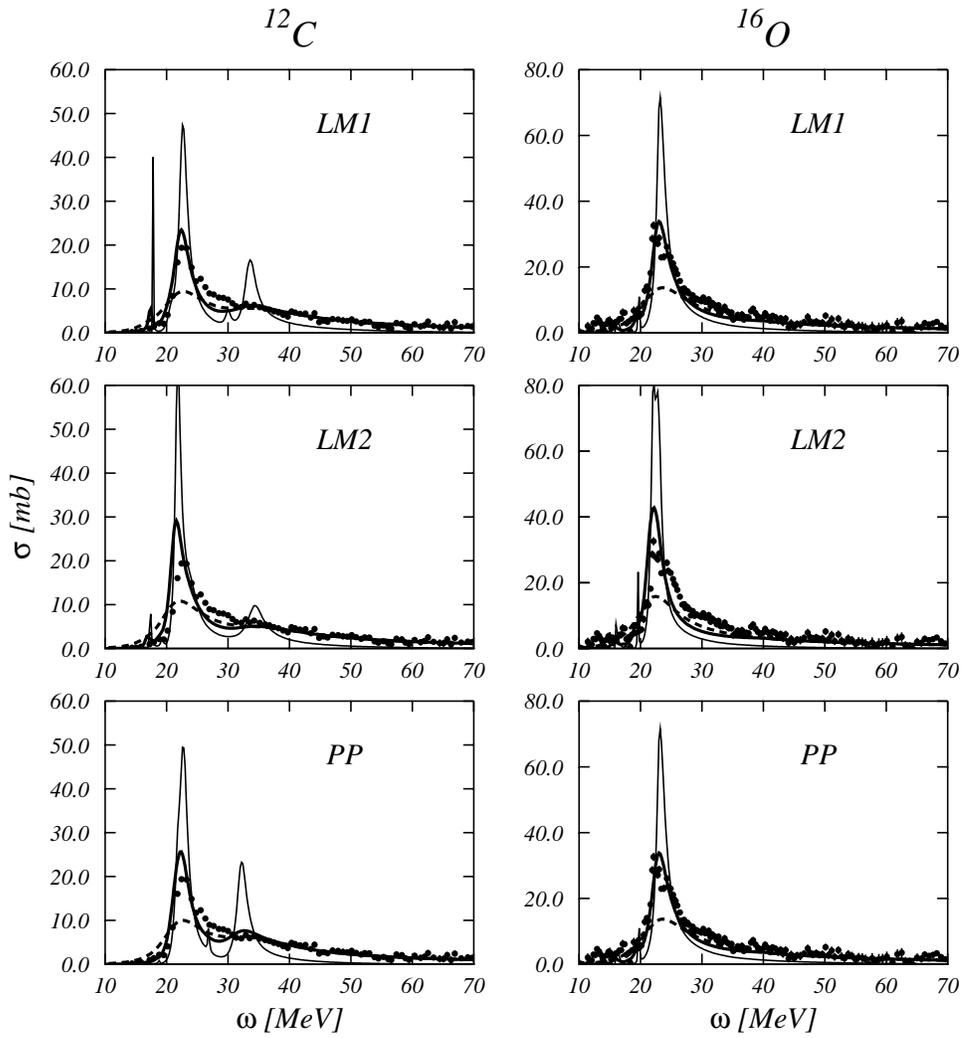}
\vskip -1.5 cm
\caption{\small The full thin lines 
  show the CRPA results, while the full thick
  lines have been obtained by applying our energy dependent FSI
  model. The data
  are from Ref. \protect\cite{ahr75} and the labels of each panel
  indicate the residual interaction used in the calculation. The dashed
  lines show the results when the parameters given in lowest row of
  Table \protect\ref{tab:gamma} are used.
 }
\label{fig:photf}
\end{figure}
\end{center}

We show in Fig. \ref{fig:photf} the effects of the model. In all the
panels of the figure the thin full lines represent the CRPA
results. The application of our FSI model by using the quasi-elastic
parameters, lowest row of Table \ref{tab:gamma}, produces the dashed
lines. The thick full lines show the results obtained with the energy
dependent parameterization.  The effect of the FSI is evident, the
peaks of the cross sections are lowered, and their widths become
larger. The asymptotic parameterization has too strong an effect on
the CRPA cross sections. In order to simplify the discussion, we have
used a unique parameterization for all the interactions and for both
nuclei, for this reason the agreement with the data is not of the same
quality for all the cases considered.

We have tested the validity of our model against the available
electron scattering data. We have not been able to find inclusive
electron scattering data in the giant resonance region for the \oxy
nucleus.  The agreement with the available \oxy quasi-elastic data
\cite{ang96} is quite satisfactory as has been shown in Ref.
\cite{co02}.

\begin{center}
\begin{figure}
\includegraphics [angle=0, scale=0.7] 
{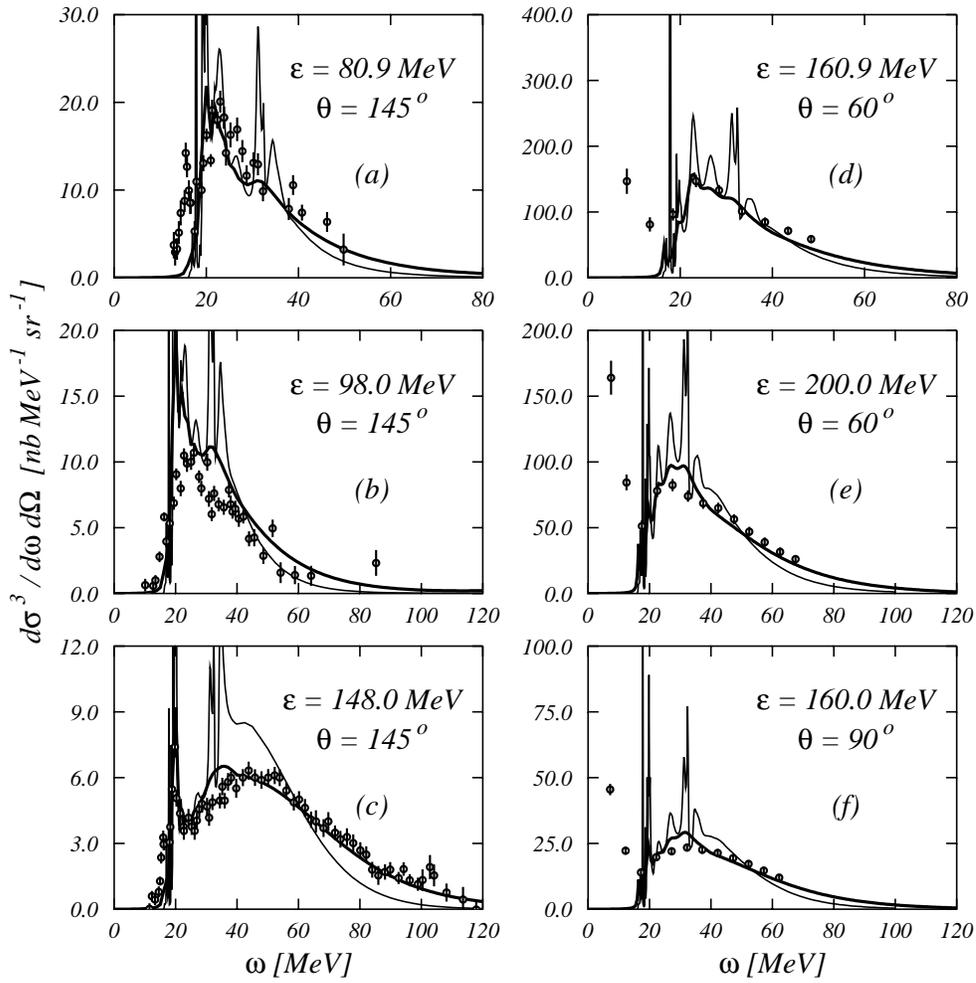}
\vskip 0.1 cm
\caption{\small The thin lines show the CRPA results, while the thicker
  lines have been obtained by applying the FSI folding model. The LM1
  interaction has been used in all the calculations. 
  The data shown in the (a), (b) and (c) panels are those quoted in Ref. 
  \protect\cite{kol97},
  while the data of the other three panels have been measured in Saclay
  \protect\cite{bar83}. 
}
\label{fig:eec12}
\end{figure}
\end{center}

The test comparison in Fig. \ref{fig:eec12} is done with \car data
only.  For the sake of brevity, we show only the results obtained with
the LM1 interaction.  The other interactions produce very similar
curves. The data shown in the (d),(e),(f) panels of the figure are
lowest energy cross sections measured in Saclay \cite{bar83} in the
framework of an experimental program aimed to separate charge and
current quasi-elastic responses by means of a Rosenbluth technique. For
this reason few points have been measured in the giant resonance
region. To the best of our knowledge, the data shown in the (a), (b)
and (c) panels have never been published. We have taken them from Ref.
\cite{kol97} where they are quoted. These data cover part of the giant
resonance region.

The results of Fig. \ref{fig:eec12} show that the application of the
FSI model improves considerably the agreement with the data. This was
somehow expected, since the set of multipoles dominating the
photo-absorption cross section and the electron scattering cross
section are quite similar. Our FSI model describes rather well the
inclusive \car electron scattering data in the quasi-elastic region
\cite{co88,ama01,ama94}.

\section{Results}
\label{sec:res}
The model described in the previous section has been applied to the
description of neutrino scattering off the \car and \oxy nuclei.  We
first discuss some characteristics of the neutrino cross sections
independent of the nuclear structure details. Then, we make a
comparison between electron and neutrino cross sections under the same
kinematic conditions. The discussion of these two points is necessary
for a better understanding of the following section where we discuss
the interplay between the nuclear structure model and the neutrino
induced nuclear excitations.  Finally, we apply our model to two
specific cases: the neutrino emission from muon decay at rest, and
from supernova explosion.

\subsection{Some general features of the neutrino-nucleus cross
  sections} 
\label{ssec:charnu}
All the results presented in this section have been obtained with CRPA
calculations by using the LM1 interaction and without the application
of the FSI model.

\begin{center}
\begin{figure}
\includegraphics [angle=0, scale=0.6] 
{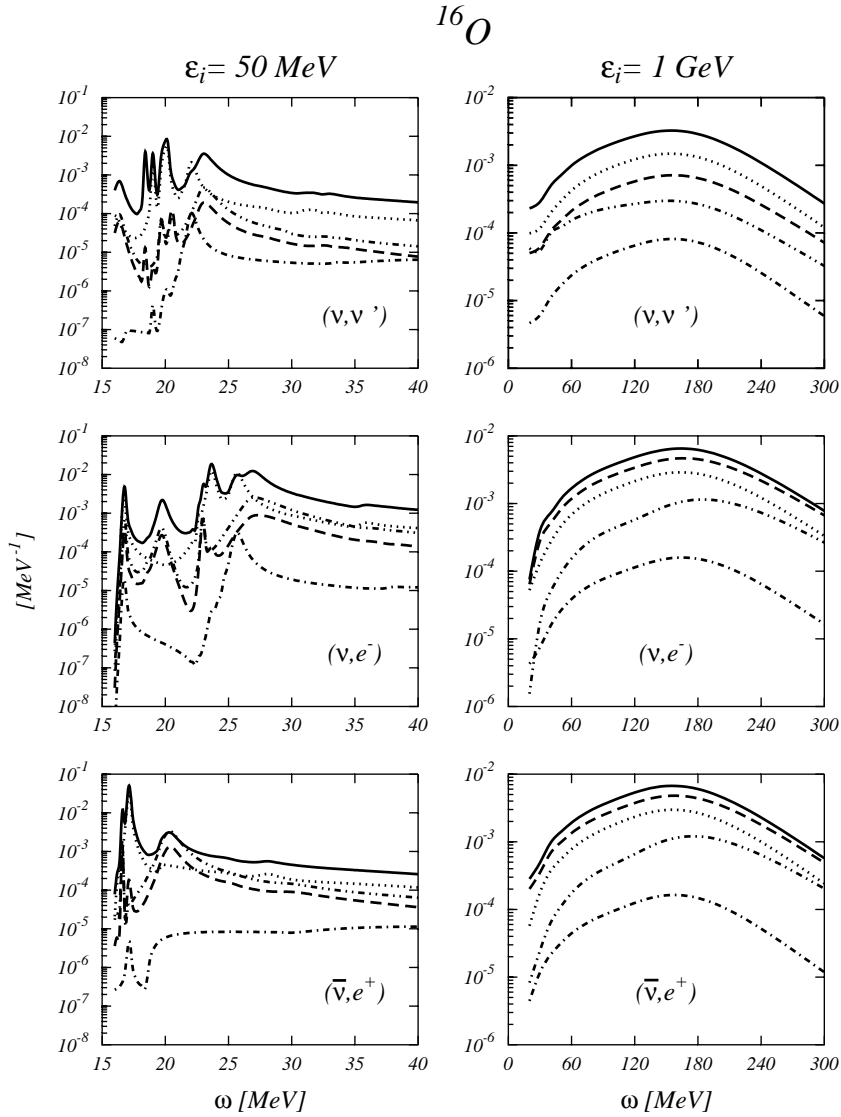}
\vskip 0.1 cm
\caption{\small Weak responses 
 calculated for $\theta=30^o$ and for two
 different values of the lepton energies.
 The full lines show the transverse axial vector
 responses. See the text, section \protect\ref{ssec:charnu},
 for the meaning of the other lines.
}
\label{fig:resp}
\end{figure}
\end{center}

In Fig. \ref{fig:resp} we show the squares of the hadronic transition
matrix elements as a function of the nuclear excitation energy.  We
have considered three neutrino scattering processes, and two neutrino
incoming energies. The transition amplitudes have been calculated for
the scattering angle $\theta$ = 30$^o$. The tree left panels show the
results for a nuclear excitation in the giant resonance region, with
momentum transfer values varying between 27 and 42 MeV/c. The results
shown in the right panels are related to the excitation of the
quasi-elastic peak with momentum transfer values varying between 512
and 526 MeV/c.

In all the panels the full lines show the contribution of the
transverse axial vector operators, while the dashed lines show the
contribution of the transverse vector operators.  The other
lines are related to the longitudinal terms of the cross section.
Specifically, the dotted lines show the axial longitudinal
contributions, the dotted-dashed those of the axial Coulomb-like, and
the dotted-dashed lines those of the vector Coulomb-like terms.

The cross section is obtained by adding to the terms shown in Fig.
\ref{fig:resp} the interference terms, and by multiplying all of them
with the leptonic terms. This changes only slightly the relative
weight of the various terms. In all the cases we have investigated,
the transverse axial vector currents dominate the cross sections
\cite{bot04}.

The other very general trend, is that of the axial Coulomb term
whose contribution to the cross sections is
always orders of magnitudes smaller than those of the other currents.
These results are in agreement with those of Ref. \cite{jac99}.  The
behavior of the other current operators is more complicated.  Their
role changes depending upon the reaction and the kinematics. For
example, the charge-exchange reactions in the quasi-elastic region
show remarkable contributions of the transverse vector operator,
usually smaller than the contributions of the longitudinal axial 
operators.

\begin{center}
\begin{figure}
\includegraphics [angle=0, scale=0.6] 
{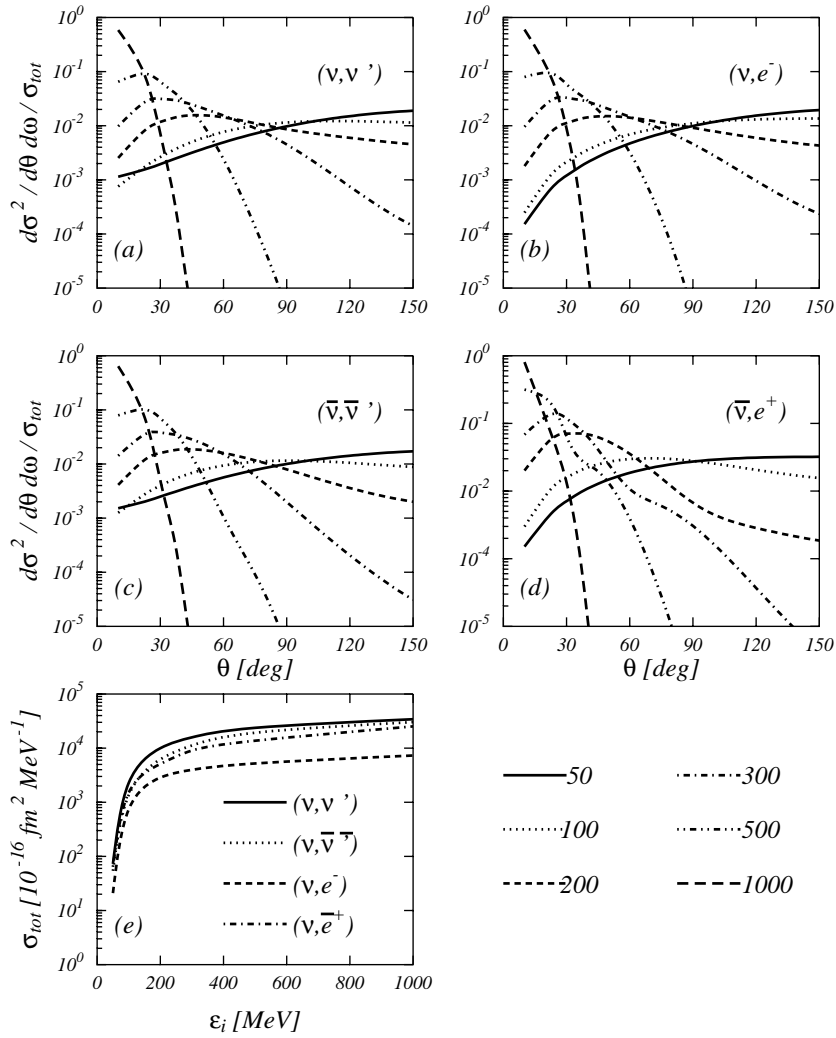}
\vskip 0.1 cm
\caption{\small 
  The $(a)-(d)$ panels show the ratios between the doubly differential
  cross sections and the angular integrated cross sections
  $\sigma_{tot}$ as a function of the scattering angle $\theta$. The
  target nucleus is \oxy.  The excitation energy for all the
  calculations is 20 MeV. The various lines have been obtained by
  changing  $\varepsilon_i$ ,the projectile energy, whose values in
  MeV are indicated in the lower part of the figure. In the panel
  $(e)$ we show $\sigma_{tot}$ as a function of $\varepsilon_i$.
}
\label{fig:ang}
\end{figure}
\end{center}

Another point we want to discuss is the angular distribution of the
emitted lepton. In Fig. \ref{fig:ang} we show the results of
calculations done for fixed nuclear excitation energy, 20 MeV, and for
different values of the projectile energy $\varepsilon_i$.  The double
differential cross sections have been integrated on the angle. These
are what we call $\sigma_{tot}$ in the figure, and their behavior as
a function of $\varepsilon_i$ is shown in the panel $e$. The ratios
between the double differential cross sections and $\sigma_{tot}$ are
shown in the panels $(a-d)$ as a function of the scattering angle
$\theta$. In these panels the various lines show the results obtained
with the values of $\varepsilon_i$ given in the lower right part of the
figure.  In agreement with the results of Refs. \cite{hax87} and
\cite{kol03} we found a common trend of all the reactions considered:
with increasing $\varepsilon_i$ the angular distributions
become forward peaked.  For the lowest values of $\varepsilon_i$ we have
used, 50 and 100 MeV, the cross sections are larger in the backward
than in the forward direction.  This behavior is more pronounced in
the charge-exchange reactions, $(b)$ and $(d)$ panels. In any case,
the relative differences between the cross section values in the
angular distributions is smaller at lower energies than at high
energies.  The absolute values of the cross sections can be estimated
from the results shown in the $(e)$ panel. The values of
$\sigma_{tot}$ increase about two order of magnitude when
$\varepsilon_i$ varies from 50 up to 200 MeV. After that, the increase of
the cross section is much slower. The results we have shown are
relative to a specific value of the excitation energy, but we found
the same behavior for all the energies investigated \cite{bot04}.

\subsection{Electron versus neutrino scattering}
\label{ssec:evsnu}
\begin{center}
\begin{figure}
\includegraphics [angle=90, scale=0.6] 
{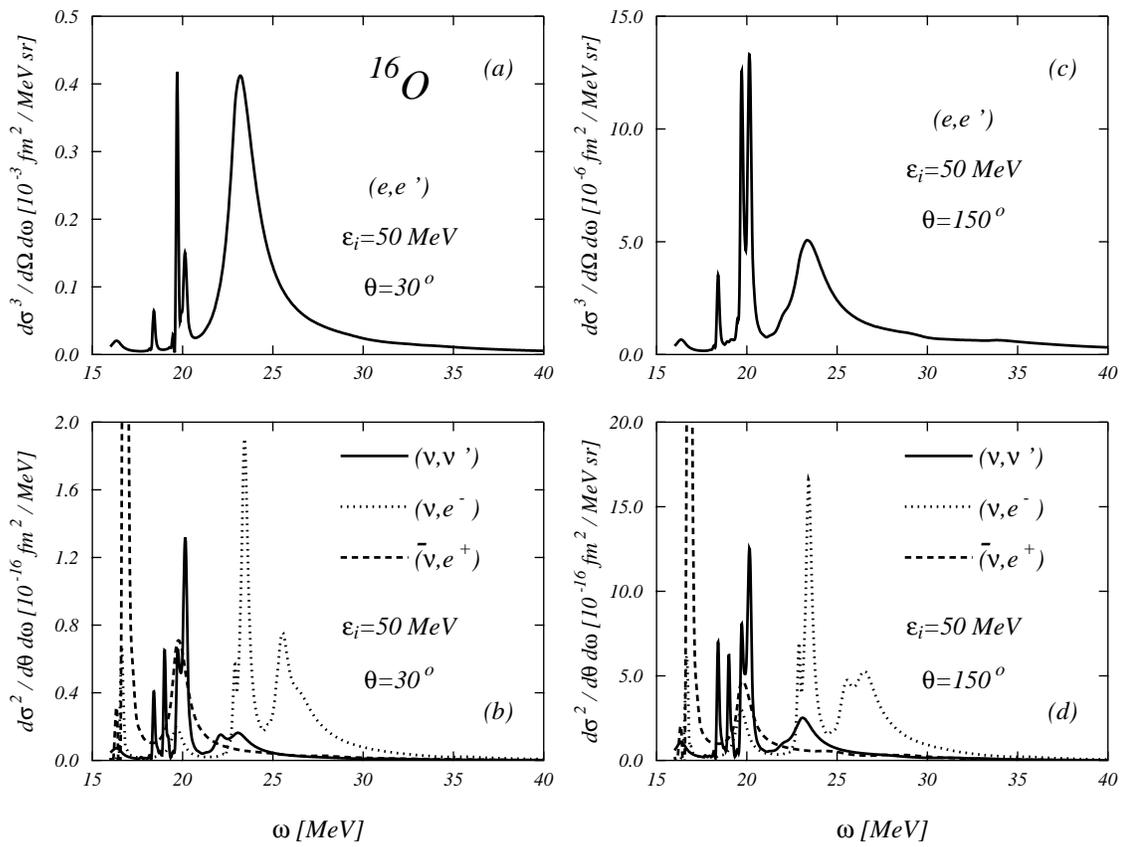}
\vskip 0.1 cm
\caption{\small Double differential cross sections for electron 
  and neutrino scattering processes, as a function of the
  nuclear excitation energy $\omega$.
  The calculations have been done in CRPA with the LM1 interaction. 
  Multipoles up to J=12 have been considered. 
  In each panel, the incoming energy $\varepsilon_i$ and the scattering
  angle $\theta$, are specified. The $10^{-16}$ fm$^2$
  units correspond to the more commonly used $10^{-42}$ cm$^2$ units. 
}
\label{fig:comp}
\end{figure}
\end{center}

\begin{center}
\begin{figure}
\includegraphics [angle=90, scale=0.6] 
{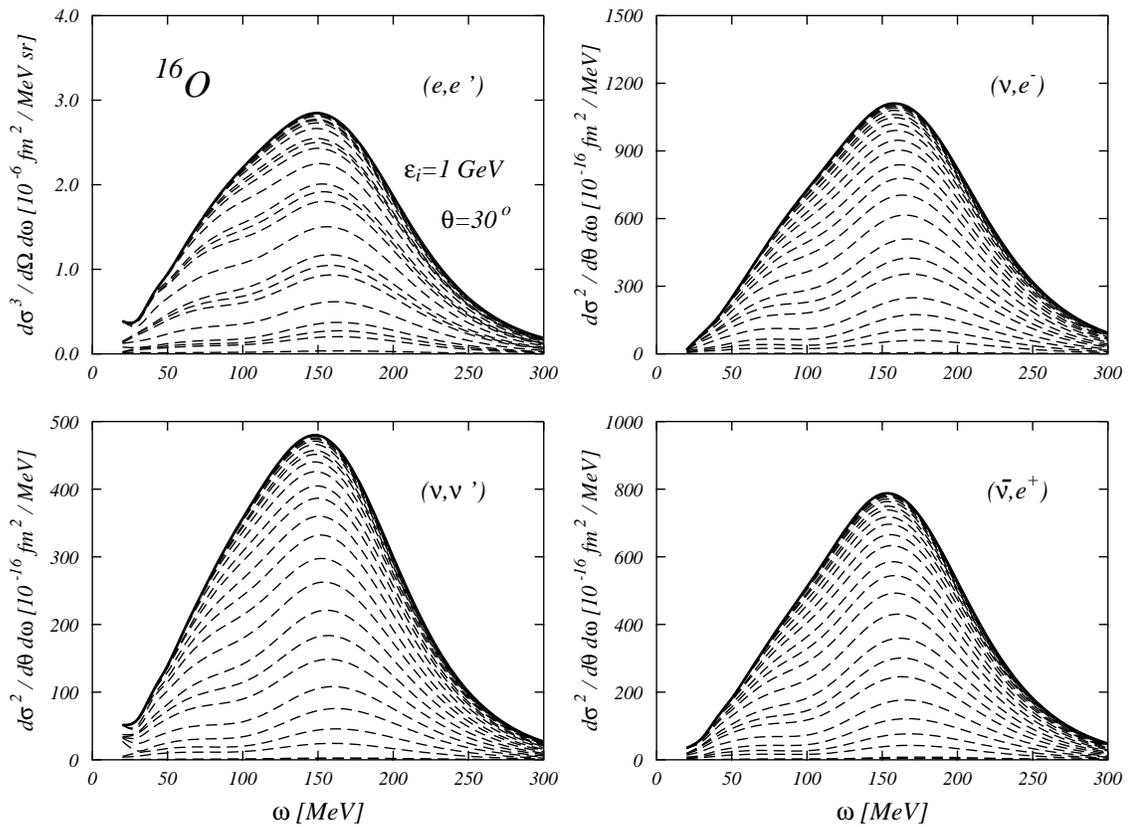}
\vskip 0.1 cm
\caption{\small The same as in Fig. \protect\ref{fig:comp} in the
  quasi-elastic region. The thin broken lines indicate the
  contribution of 
  each multipole excitation to the total cross sections represented by
  the thick full lines. 
  These contributions are summed one on top to the other ones, and they
  are ordered with increasing value of angular momentum J.
  For a value of J, we give first the contribution of 
  the negative parity state, and after that of positive parity. 
}
\label{fig:qecomp}
\end{figure}
\end{center}

There are many analogies between electron and neutrino scattering
processes.  For this reason electron scattering data can be used to
make predictions about neutrino cross sections \cite{lan04,ama05}.  In
Figs.  \ref{fig:comp} and \ref{fig:qecomp} we compare the nuclear
excitations induced by both probes, in the same kinematic conditions.
All the results have been obtained by using the CRPA and considering
multipole excitations up to J=12.  In Fig.  \ref{fig:comp} the energy
of the incoming lepton has been fixed at 50 MeV, and two values of the
scattering angle have been used: $\theta$ = 30$^o$ and 150$^o$.  In
Fig.  \ref{fig:qecomp} we considered a lepton incoming energy of 1 GeV
and $\theta$ = 30$^o$.

Let's first discuss Fig. \ref{fig:comp} where the nuclear giant
resonances are in general excited.  The behavior of the electron
scattering cross sections is well known.  The cross section at forward
scattering angles is ruled by the longitudinal response, which is
excited only by natural parity multipoles.  At backward angles the
transverse response, excited by both natural and unnatural parity
multipoles, dominates.  Furthermore, the values of the cross sections
decrease with increasing incoming energy and scattering angle.  All
these characteristics are present in our results.  It is worth
pointing out the change of scale in the panels $(a)$ and $(c)$.  To
facilitate the discussion we show in Table \ref{tab:mul} the percentage
contribution of the various multipoles to the energy integrated cross
sections of Fig. \ref{fig:comp}.  The electron scattering cross
section is dominated by the 1$^-$ excitation for $\theta$=30$^o$, as
is shown in Table  \ref{tab:mul}. The same table shows that, with
increasing scattering angle, the role of the 1$^-$ excitation
decreases, while that of the 2$^+$ and of the 2$^-$ increases. This is
a sign of the relevance of the transverse response.  The broad peaks
at 23.2 MeV are due to the 1$^-$, multipole responsible also of the
sharp peak at 19.7 MeV. The sharp excitation at 20.1 MeV, which is
enhanced at $\theta$=150$^o$, is due to the 2$^-$ multipole.

The situation is quite different for the neutrino scattering cross
sections.  The shapes of all the neutrino cross sections have little
resemblance with those of electron scattering cross section.  A
difference with the charge-exchange reactions was expected, since the
basic particle-hole transitions are not those excited in electron
scattering. The difference with the charge-conserving cross section is
more surprising.  It is shown in Table \ref{tab:mul} that the neutrino
cross sections are not dominated by the 1$^-$ multipole, even at small
scattering angles. In effect, the contribution of the 2$^-$ multipole
is relevant in all the cross sections. This multipole is 
responsible for the peak at 20.1 MeV in the $(\nu,\nu\,')$ cross
section.  We have also verified that  the $(\overline {\nu}, \,
\overline{\nu}\,')$ cross section, which is not shown in the figure,
has a sharp peak in the same position, corresponding to the analogous
excitation in $(e,e\,')$.  It is interesting to notice that in the
experimental electromagnetic spectrum, there is a 2$^-$ state at 20.43
MeV \cite{led78}.  Our $(\nu,e^-)$ cross section shows a 2$^-$ sharp
peak at 23.40 MeV to be compared with a 2$^-$ state at 24.09 MeV found
in $(p,n)$ reactions \cite{faz82}.
 
The features we have just discussed are understood by observing that,
in all the neutrino processes we have studied, the axial vector part
of the weak current dominates on all the other cross section terms.
To be more specific, as shown in Fig. \ref{fig:resp}, the transverse
axial terms, are usually orders of magnitude larger than the other
terms of the cross section. For this reason, in neutrino scattering
there is no dominance of the natural parity multipoles, contrary to
what happens with the electromagnetic interaction.

The dominance of the axial current has consequences also in the
quasi-elastic excitation, as we show in Fig. \ref{fig:qecomp}.  The
shapes of the various cross sections are rather similar, and they show
the quasi-elastic peak at the same value of the excitation energy
$\omega$ = 150 MeV. The differences become evident when the
contribution of each multipole to the total cross section is
explicitly studied.  In the figure the progressive sum of these
contributions, which finally gives the full line, is presented by the
thin dashed lines.  The contributions are ordered with increasing
value of the angular momentum, and for each multipole, we show first
the negative parity contribution, and then, that of the positive
parity.  The staggering shown in the electron scattering results, is
due to the fact that natural parity multipoles contribute more than
the unnatural parity ones, because the relatively small value of the
scattering angle favors the longitudinal response. In neutrino
scattering, the contribution of the various multipoles is more
regularly distributed, without any particular difference between
natural and unnatural parity states.  From the above observations we
conclude that the procedure proposed in Ref. \cite{ama05} to predict
neutrino cross sections by using universal response functions
extracted from electron scattering data, should be used with caution,
if at all.

An analogous study has been done for the \car nucleus. For sake
of brevity we do not show here the results, which in the case
of quasi-elastic region are similar to those of \oxy.  In the giant 
resonance region, we observed that in \car, the role of the 2$^-$ is
not as remarkable as in \oxy. In \car the excitation of the 1$^+$
multipole is more important.

\subsection{The sensitivity to the nuclear model}
\label{ssec:totcs}

In this Section we discuss the stability of our results against the
uncertainties related to the choice of the residual interaction, and
the influence of the FSI.

\begin{center}
\begin{figure}
\includegraphics [angle=90, scale=0.6] 
{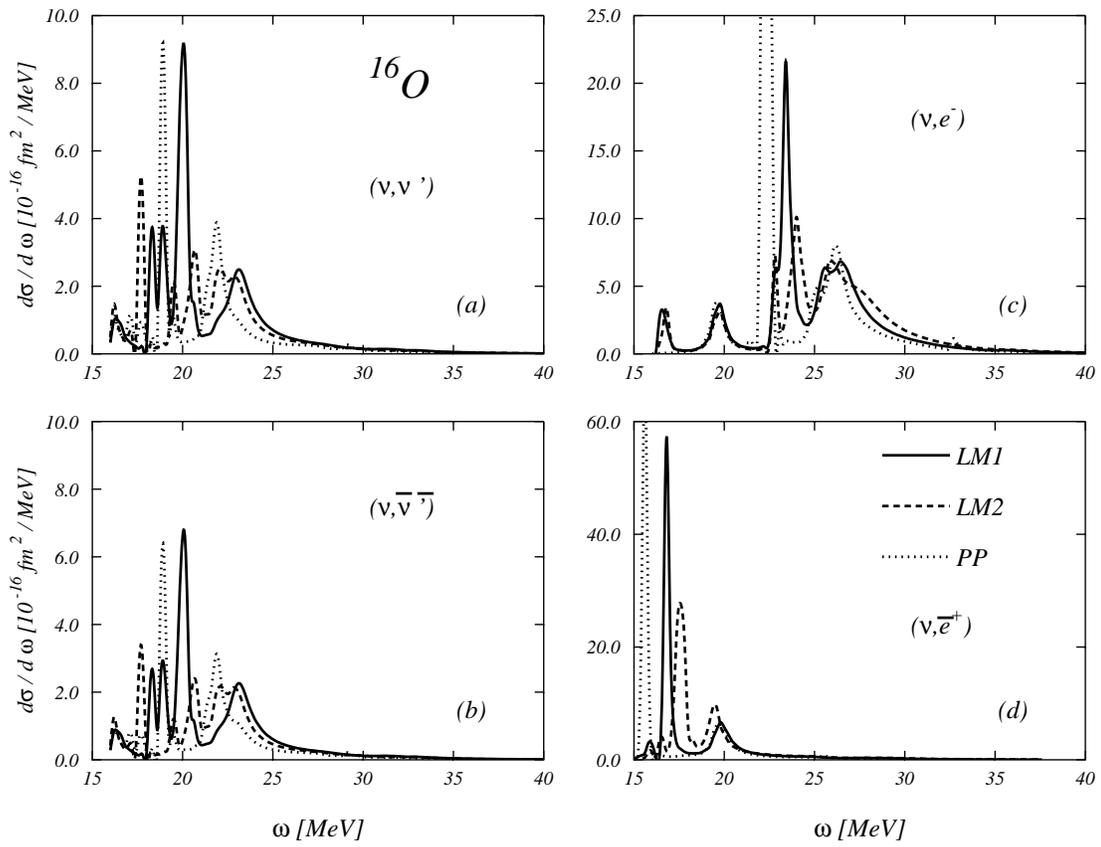}
\vskip 0.1 cm
\caption{\small Differential cross sections on the \oxy nucleus 
 for various neutrino reactions
 integrated on the angular distributions, as a function of the
 nuclear excitation energies. In all the cases the values of the projectile
 energy, $\varepsilon_i$, is 50 MeV. The various cross sections
 have been calculated with the CRPA by using the three
 interactions adopted in this paper. The meaning of the lines is given
 in the panel (d).
}
\label{fig:LMPPwam}
\end{figure}
\end{center}

In Fig. \ref{fig:LMPPwam} we show the differential cross sections for
various neutrino scattering processes. The target is the \oxy. The
projectile energy, $\varepsilon_i$, has been fixed at 50 MeV, and the
double differential cross sections have been integrated on the angular
distribution of the scattered lepton.  The cross sections have been
obtained by using the three residual interactions presented in Sect.
\ref{ssec:crpa}.  The dispersion of the results of Fig.
\ref{fig:LMPPwam} is larger than that relative to the photo-absorption,
see Fig. \ref{fig:sr}. The positions of the peaks change with the
interaction, and also the heights and the widths of the resonances.
This sensitivity to the residual interaction is comparable to that
shown in Refs. \cite{jac99} and \cite{kol92}.

In Fig. \ref{fig:LMPPwam} the cross sections show narrow peaks.  In
our CRPA calculations the widths of the resonances are due,
exclusively, to the proper treatment of the continuum, the so-called
escape width. The various processes shows different type of
resonances.  The $(\anu,e^+)$ cross sections shows peaks between 15
and 20 MeV, while the $(\nu,e^-)$ cross sections between 22 and 30
MeV.  The resonances of the charge-conserving $(\nu,\nu ')$ and
$(\anu,\anu ')$ reactions are positioned between 19 and 25 MeV.  Also the
widths of these peaks are quite different. They are narrower in
$(\anu,e^+)$ than in the other processes. All these features can be
understood by considering that the different processes excite
different $p-h$ configurations.

The comparison between the two charge-conserving reactions, panels
$(a)$ and $(b)$ of Fig. \ref{fig:LMPPwam}, indicates that, in our
calculations, neutrinos and antineutrinos excite the same resonances.
For each interaction, our results show a perfect coincidence between
the position of the peaks of the neutrino and those of the
antineutrinos cross sections. This was an expected result, since the
two processes imply the same configuration space.  We did not find any
trace of the effects presented in Refs. \cite{kol92} and
\cite{kol92a}.

In general, the antineutrino cross sections are smaller than the
neutrino ones.  This effect becomes evident in the total cross
sections.  

\begin{center}
\begin{figure}
\includegraphics [angle=90, scale=0.6] 
{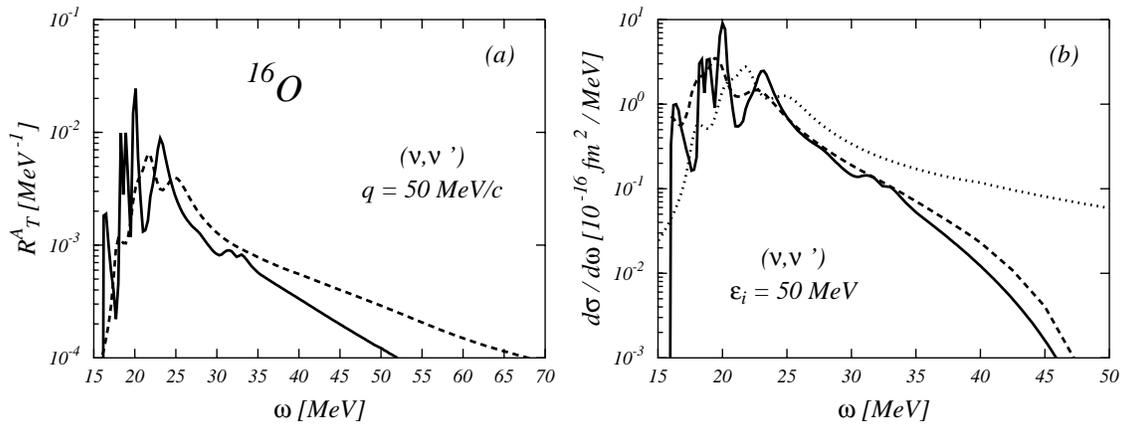}
\vskip 0.1 cm
\caption{\small In panel (a) the \oxy 
  transverse axial response at $q$ = 50
  MeV/c for the $(\nu,\nu')$ process is shown as a function of the
  nuclear excitation energy. The full line show the CRPA result
  obtained with the LM1 interaction,
  the dashed line show the result of the folding procedure.
  In panel (b) the angular integrated cross sections for the 
  $(\nu,\nu')$ process are shown as a function of the excitation
  energy. The full line shows the CRPA result. The dotted line is the
  result obtained when the folding procedure is applied to the cross
  section. The dashed line has been obtained by applying the folding
  procedure to the responses.
}
\label{fig:fold}
\end{figure}
\end{center}

All the cross sections shown in Fig. \ref{fig:LMPPwam} have been
calculated within the CRPA framework.  To illustrate the effects of
the FSI, we show in the panel (a) of Fig. \ref{fig:fold} the
transverse axial response.
The full line shows the CRPA response obtained with the LM1
interaction. The value of the momentum transfer has been fixed, and
the response has been calculated for various values of the nuclear
excitation energy.  It is important to notice that, in the response,
the two kinematic variables $q$ and $\omega$ are independent.  The
application of our FSI model, produces the dashed line. As expected,
the CRPA strength is redistributed. The peaks become smaller and
broader, and the strength is mainly shifted toward the high energy
tail.  The strength moved at higher energies is lost when the response
is inserted in the cross section, since in this case $q$ and $\omega$
are related by kinematics constraints. For example $\omega$ cannot be
larger than $q$. This fact provokes a general reduction of the total
cross section.

In the panel (b) of Fig. \ref{fig:fold} we show the effect of the FSI
on the cross section. The full line shows the CRPA result. The naive
application of the folding model to the cross section produces the
dotted line. This procedure is incorrect, since it involves also the
leptonic terms, which are momentum and energy dependent.  The correct
procedure consists in applying the folding model to the responses.
The cross section is then calculated by using the folded responses.
In the figure, the dashed line shows the cross section obtained in
this last, correct, manner.  Also in this case the peaks are lowered
and smoothed, but their positions do not change as strongly as in the
naive approach.  More striking is the fact that the behavior in the
higher energy region is rather similar to that of the unfolded cross
section.

\begin{center}
\begin{figure}
\includegraphics [angle=0, scale=0.6] 
{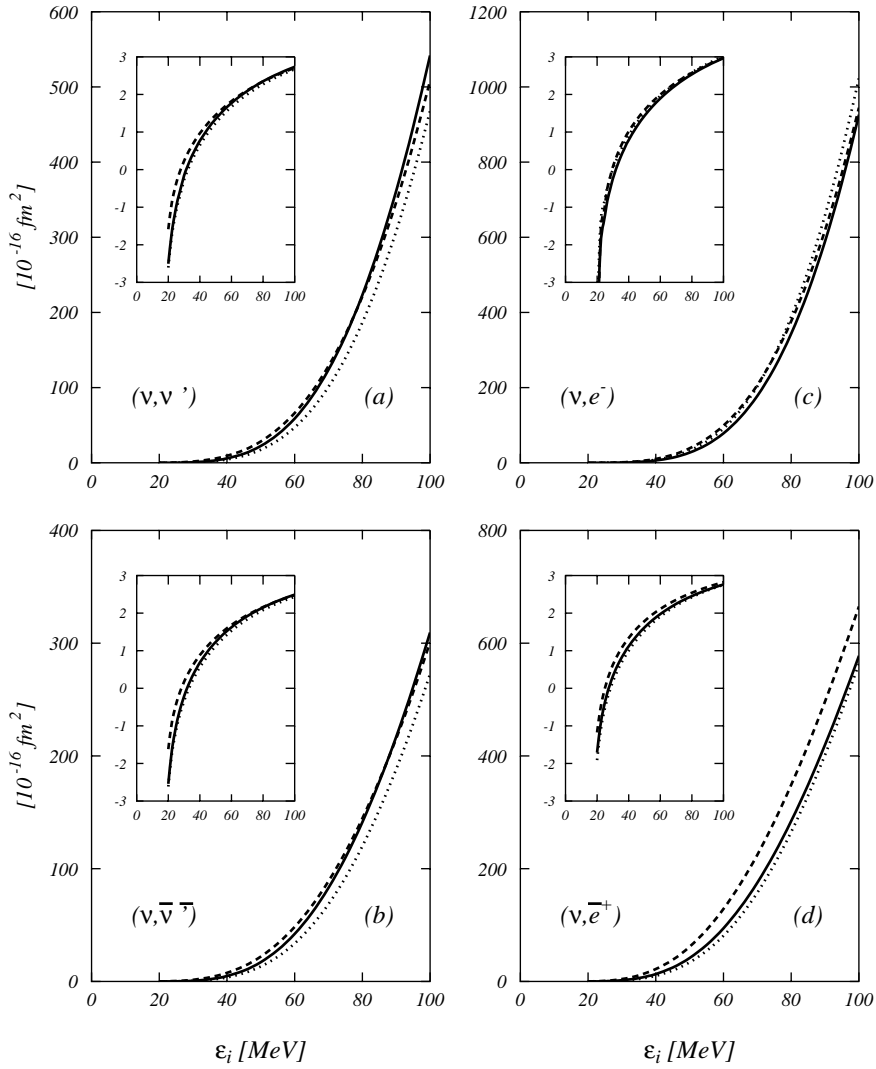}
\vskip 0.1 cm
\caption{\small Total cross sections on a \car target 
  as a function of the projectile
  energy $\varepsilon_i$. The full lines show the LM1 results,
  the dashed lines the LM2 results and the dotted lines the PP
  results. In the inserts, the same curves are shown in logarithmic
  scale. The numbers on the y axes indicates the powers 10 of the 
  logarithm. 
}
\label{fig:totc12}
\end{figure}
\end{center}

\begin{center}
\begin{figure}
\includegraphics [angle=0, scale=0.6] 
{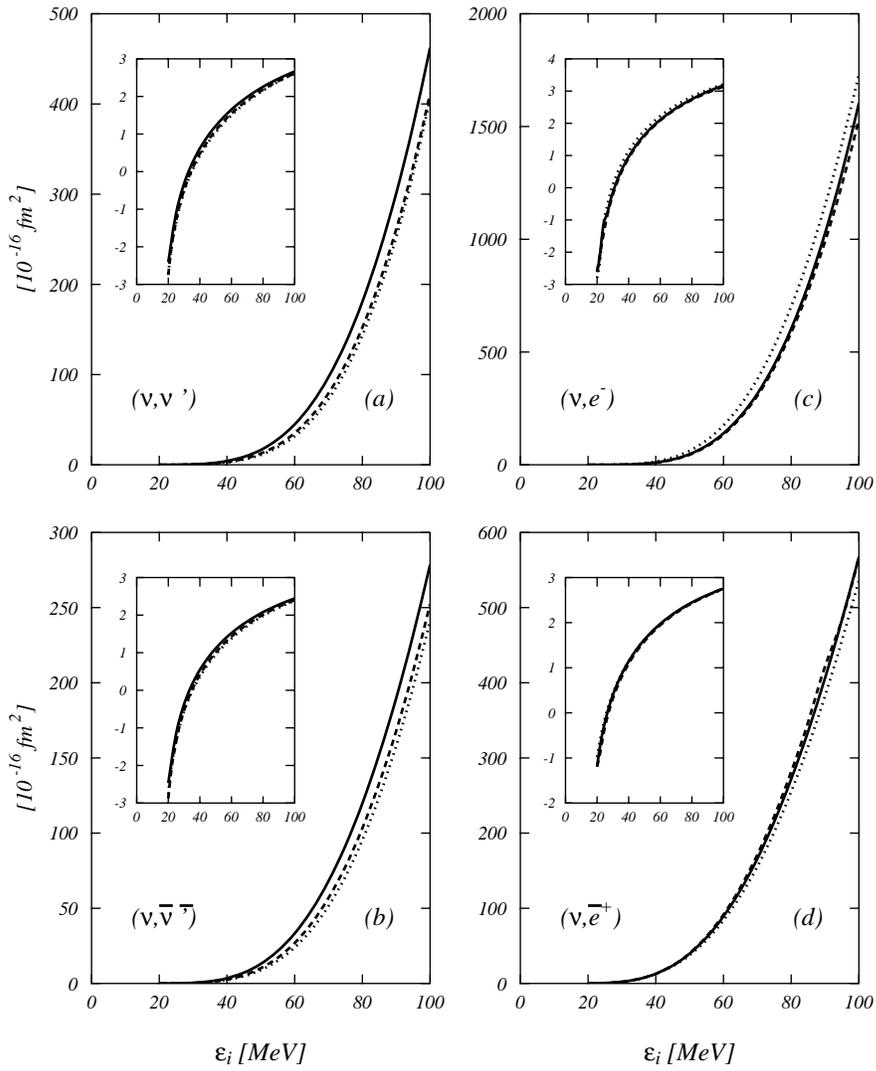}
\vskip 0.1 cm
\caption{\small  The same as Fig. \protect\ref{fig:totc12} for the
  \oxy nucleus.
}
\label{fig:toto16}
\end{figure}
\end{center}

The total cross sections, obtained by integrating on the scattering
angle and on the nuclear excitation energy, are shown in Figs.
\ref{fig:totc12} and \ref{fig:toto16}, for \car and \oxy respectively,
as a function of the projectile energy $\varepsilon_i$.  Since the cross
sections values vary by orders of magnitude, we present them in both
linear and logarithmic scales, which emphasize different aspects.
The cross sections shown in the figures have been obtained by using
the three interactions, and all contain the FSI effects. These are the
main results of our work.

\begin{center}
\begin{figure}
\includegraphics [angle=0, scale=0.6] 
{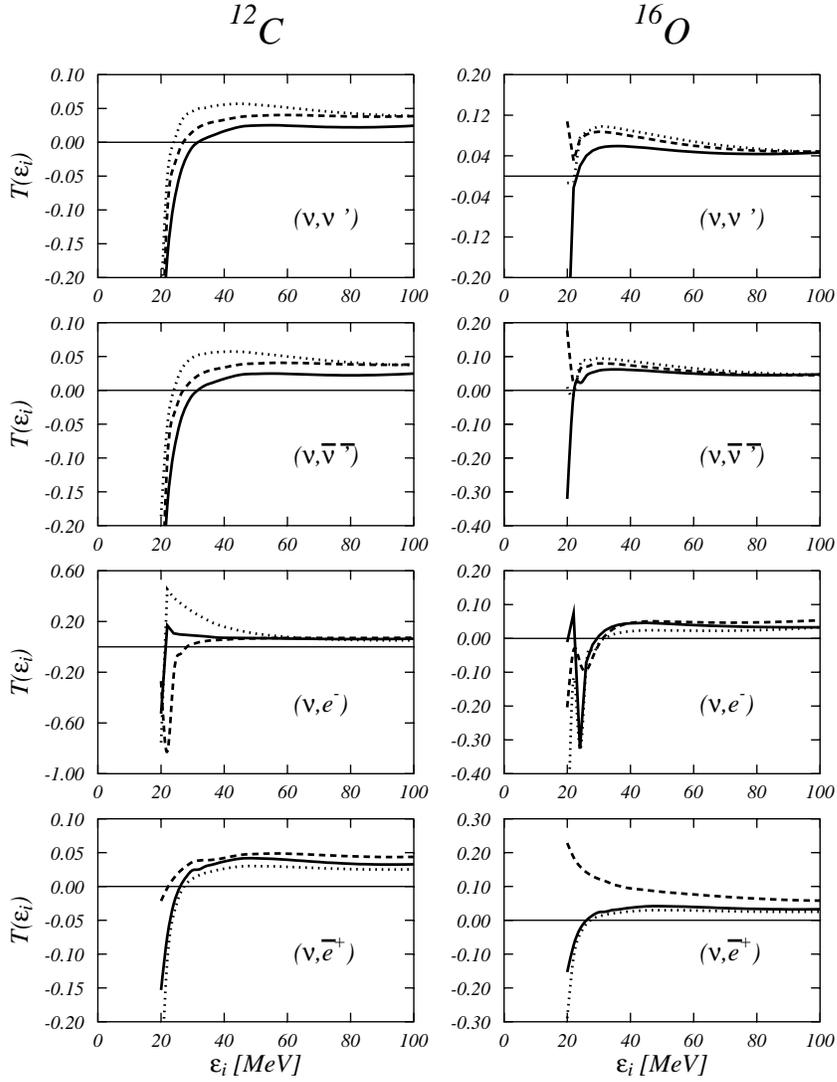}
\vskip 0.1 cm
\caption{\small Ratio $T(\varepsilon_i)$, Eq. (\protect\ref{eq:ratio}),
  for the four neutrino interactions
  considered in this work. As in the previous figures, the full lines
  show the LM1 results, the dashed lines the LM2 results and the
  dotted lines the PP results.
}
\label{fig:ratio}
\end{figure}
\end{center}

In Fig. \ref{fig:ratio} we present the effects of the FSI 
by showing the following ratios between total cross sections:
\begin{equation}
T(\varepsilon_i) = 
\frac {\sigma^{RPA}(\varepsilon_i) - \sigma^{FSI}(\varepsilon_i)}
      {\sigma^{RPA}(\varepsilon_i) + \sigma^{FSI}(\varepsilon_i)}
\label{eq:ratio}
\end{equation}
as a function of the projectile $\varepsilon_i$. for the various total
cross sections. The behavior of $T(\varepsilon_i)$ at low values of
$\varepsilon_i$, can be understood by considering that, in the evaluation
of the total cross sections,  only the differential cross
sections with excitation energies close to the nucleon emission
threshold enter. Part of the strength removed by the FSI from the giant
resonance region, goes to lower energies, i.e. in the region we are
discussing. Therefore, in general, at low $\varepsilon_i$ the FSI cross
section is larger than $\sigma^{RPA}$. The exceptions to this trend
  are due to the presence of sharp peaks in the threshold region.
  These peaks are smoothed by the FSI. At low $\varepsilon_i$ values the
  sensitivity to the residual interaction is remarkable.

This situation changes at higher  $\varepsilon_i$ values, where the
various results stabilize their behaviors around a specific
value. For  $\varepsilon_i \ge $ 40 MeV, the effect of the FSI presented,
and discussed,  in Fig. \ref{fig:fold} starts to be effective,
therefore the  $\sigma^{FSI}$ is smaller than  $\sigma^{RPA}$, and
this produces the positive values of $T(\varepsilon_i)$.
The
asymptotic values of the ratios indicate the reduction factor produced
by the FSI. These values vary between 0.05 and 0.10 in agreement with
the results of Refs. \cite{co02,ble01,bot05} where the
quasi-elastic regions have been investigated.

The effects of the various residual interactions on the total cross
sections of Figs. \ref{fig:totc12} and \ref{fig:toto16} do not have
systematic trends. Each reaction, and nucleus, should be separately
analyzed.  The logarithmic plots indicate that, at low energy values,
the differences in \car are larger than those found in \oxy. It is
evident that LM2 results in the \car $(\anu,e^+)^{12}B$ reactions are
remarkably larger than the other ones.  This is due to the fact that
in the LM2-CRPA calculation a strong 1$^+$ resonance appears above the
nucleon emission threshold. This does not happens for the other two
interactions.

\begin{center}
\begin{figure}
\includegraphics [angle=0, scale=0.6] 
{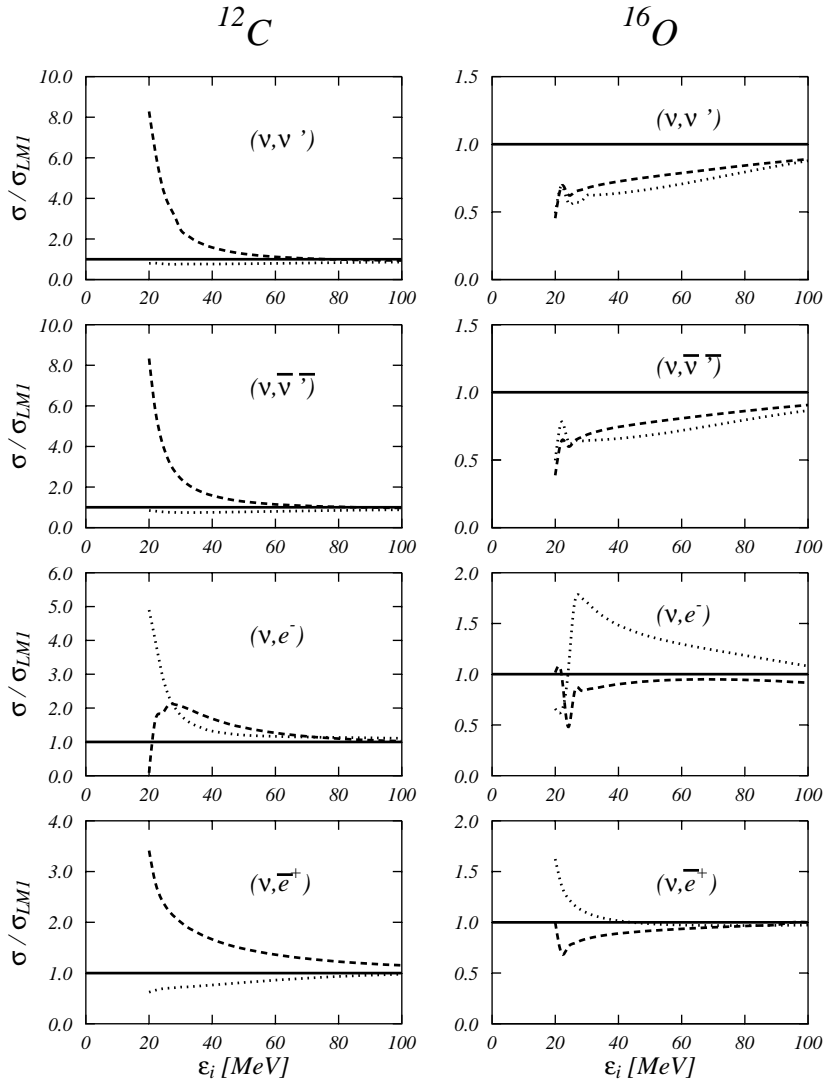}
\vskip 0.1 cm
\caption{\small Ratios between the total cross sections of Figs.
  \protect\ref{fig:totc12} and \protect\ref{fig:toto16} 
  and the LM1 cross sections. The dotted and dashed lines show the PP and
  LM2 ratios, the full horizontal line with value 1, refers to the LM1
  results.
}
\label{fig:other}
\end{figure}
\end{center}

We emphasize the differences between the total cross sections
calculated with the various interactions, by showing in Fig.
\ref{fig:other} the ratios between the total cross sections and the
LM1 total cross sections taken as a reference calculation.  The
meaning of the lines is the same as in Figs. \ref{fig:totc12} and
\ref{fig:toto16}, therefore the LM1 results are represented by
horizontal lines with value 1.  In \car the LM2 results are larger
than those obtained with the LM1 interaction for all the reactions but
for the $(\nu,e^-)$ one.  In this case, the PP result is noticeably
different from the other two.  At higher energies there is a
convergence of all the results. As we have already pointed out the LM2
results for the $(\anu,e^+)$ reaction, remain above the other ones,
even in the asymptotic region.

The results for the \oxy nucleus show a much better agreement, or in
other words, a smaller sensitivity to the residual interactions.  As
in \car the main differences between the various calculations are
located in the low energy region.  All the calculations show
convergence at high energies.  Only the results of the
charge-conserving reactions at 100 MeV do not converge to LM1 values.

\subsection{Specific applications}
\label{ssec:appl}
We have used the total cross sections of Figs. \ref{fig:totc12} and
\ref{fig:toto16} to study two specific situations: the electron
neutrinos emitted by the muon decay at rest, and those emitted after a
supernova explosion.

\begin{center}
\begin{figure}
\includegraphics [angle=90, scale=0.6] 
{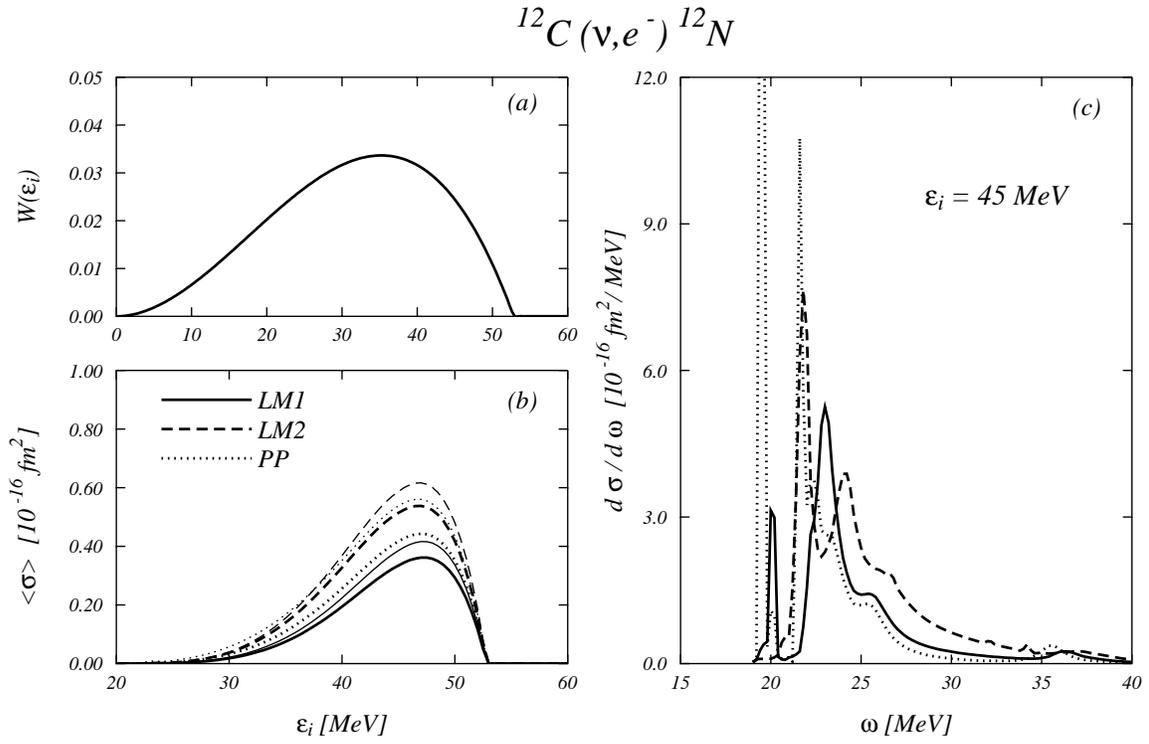}
\vskip 0.1 cm
\caption{\small In panel (a) we show 
 the normalized neutrino flux coming from the neutrino decay at rest.
 In panel (b) the flux averaged cross
 sections are shown as a function of the neutrino energy. The thin
 lines have been obtained by folding the CRPA results and the thicker
 lines with the FSI included. The different line types indicate the
 residual interactions used in the calculations. The CRPA 
 differential cross
 sections for $\varepsilon_i$ = 45 MeV are shown in panel (c) as a
 function of the nuclear excitation energy. 
}
\label{fig:michel}
\end{figure}
\end{center}

The energy distribution of the neutrinos emitted by a muon, follows the 
law \cite{mic50}:
\begin{equation}
W(\varepsilon_i) = {\mathcal N} \varepsilon_i^2
\left(\frac{m^2_\mu - m^2_e}{2 m_\mu} - \varepsilon_i \right)
\label{eq:michel}
\end{equation}
where $m_\mu$ and $m_e$ are the muon and electron mass respectively,
and $\mathcal N$ a normalization constant fixed such that the integral
of $W(\varepsilon_i)$ is 1. The normalized function $W(\varepsilon_i)$ is
shown in the panel (a) of Fig. \ref{fig:michel}.

The \car$(\nu,e^-)^{12}N$ flux averaged cross sections 
\begin{equation}
< \sigma (\varepsilon_i) > = \sigma (\varepsilon_i)\, W(\varepsilon_i)
\label{eq:aves}
\end{equation}
are presented in the panel (b) of Fig. \ref{fig:michel} as a function
of the neutrino energy.  All the results obtained with the three
residual interactions, with and without the inclusion of the FSI, are
shown in the figure.  The thinner lines present the CRPA results,
while the thicker lines are the results which include the FSI. The results
obtained with the same residual interaction are identified by the same
line type.  The FSI interaction lowers the CRPA results in all the
cases. This effect is larger for the PP potential than in the other
two cases.  At the peak of the distribution the lowering effect is
22\% for the PP and 12\% for the other two interactions. The effect of
the FSI is comparable with the uncertainty related to the choice of
the residual interaction. 

The maxima of the flux averaged cross sections are around
$\varepsilon_i$=45 MeV. For this value of the neutrino energy, we show in
the panel (c) of Fig. \ref{fig:michel}, the behavior of the angular
integrated differential cross sections as a function of the nuclear
excitation energy. The curves represent the CRPA results, without FSI.
The LM1 cross section show a large resonance around 24 MeV, and a
narrow peak at 20 MeV. In the PP cross section, this last peak is
positioned at lower energy, and it is higher.  The LM2 and PP cross
sections exhibit their main peaks at the same energy. The LM2
result, shows, in addition, smaller peak at about 25 MeV. These
differences are the source of the variations in the flux averaged
cross sections of panel (b).

A comparison with experimental data \cite{kra92,ath92,bod94} is done
in Table \ref{tab:cross}. In this table we show the values of the total
cross sections obtained by integrating the flux averaged cross
sections of the panel (b). Our results are lower than the empirical
ones. They are also remarkably lower than those obtained by other
authors which used ordinary RPA \cite{aue97}, Continuum RPA
\cite{kol99,kol94,jac02a}, quasi-particle RPA \cite{vol00,krm02} shell
model \cite{hay00}, infinite nuclear matter models in local density
approximation \cite{sin93,nie05}.  The source of these discrepancies is
related to the fact that our cross sections have been calculated only
above the continuum threshold.  We have neglected the excitation of
the discrete spectrum.  Some of these discrete excitations are
important for the neutrino energies involved in the process, for
example the 1$^+$ state at 17.34 MeV, just below the continuum
threshold. This can explain why calculations working with discrete
configuration spaces \cite{aue97,vol00,krm02,hay00} produce cross
sections larger than ours.

The other application of our results we want to discuss is related to
the inelastic neutrino scattering reactions relevant for the
nucleosynthesis following the supernova explosion.  Monte Carlo
simulations \cite{jan89,kei03} indicate that the energy distribution
of the neutrinos produced in a supernova explosion can be rather well
described by the function:
\begin{equation}
f(\varepsilon_i) = {\mathcal C} \, 
\frac {\varepsilon^2_i}
{1+exp(\varepsilon_i /T - \alpha)}
\label{eq:fluence}
\end{equation}
where $\mathcal C$ is a constant fixed to normalize $f(\varepsilon_i)$ to
unity. 
The average energy of the produced electron neutrinos,
\begin{equation}
< \varepsilon_i > = \int_0^\infty \, \varepsilon_i \, f(\varepsilon_i) \,
d\,\varepsilon_i
\label{eq:avene}
\end{equation}
can assume values from 11 up to 12 MeV, and the
parameter $\alpha$ can vary between 0 and 5 \cite{kei03}. 

\begin{center}
\begin{figure}
\includegraphics [angle=0, scale=0.6] 
{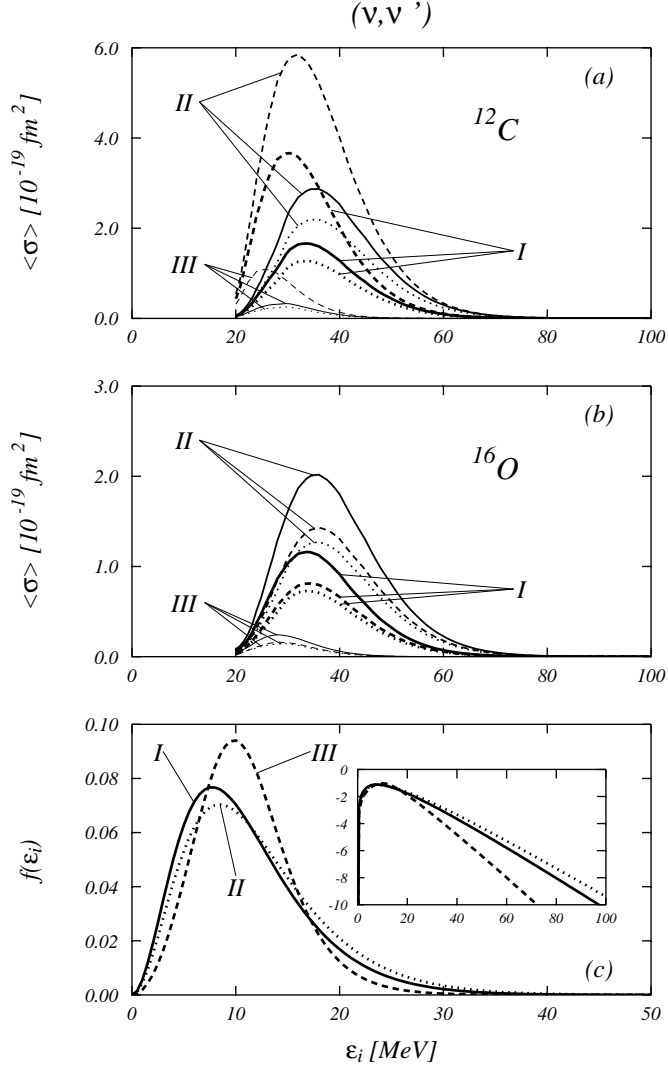}
\vskip 0.1 cm
\caption{\small Total $(\nu,\nu')$ cross sections averaged with the
  energy distribution (\protect\ref{eq:fluence}). The roman numbers
  refer to the parameters of $f(\varepsilon_i)$ given in Table
  \protect\ref{tab:fluence}. In the panels (a) and (b) the different
  line types refer to the residual interactions used in the
  calculations. As in the previous figures, the full lines indicate
  the results obtained with LM1, the dashed lines those obtained with
  LM2 and the dotted lines the results obtained with the PP. The three 
  energy distributions $f(\varepsilon_i)$ are shown in the panel (c). 
  In the insert the same lines are shown in semi-log scale.
}
\label{fig:fluence}
\end{figure}
\end{center}

We have multiplied the $(\nu,\nu')$ cross sections of Figs.
\ref{fig:totc12} and \ref{fig:toto16} with the energy distribution
(\ref{eq:fluence}). We wanted to compare the effects of the
nuclear structure uncertainties with those related 
to $f(\varepsilon_i)$. For this reason our cross sections have
been multiplied by the energy distributions (\ref{eq:avene}), obtained
with three different parameters sets, whose values are given in Table
\ref{tab:fluence}. The respective energy distributions are shown in
the panel (c) of Fig. \ref{fig:fluence}.

The flux averaged cross sections for \car and \oxy nuclei are
presented in the panels (a) and (b) of Fig. \ref{fig:fluence}
respectively.  For each nucleus we show the results obtained by using
the three different interactions for the three different
parameterizations of $f(\varepsilon)$.  The various types of line
identify the interactions used, and the roman numbers the various
parameterizations of $f(\varepsilon_i)$.

We have verified that the effect of the FSI is negligible. The
relevant nuclear structure effects are related to the residual
interactions. The uncertainties produced by different choices of the
interaction are comparable with those related to the choice of the
$f(\varepsilon_i)$ parameters.

We should point out that our cross sections have been calculated only
for $\varepsilon_i >$ 20 MeV, and for nuclear excitation energies above
the continuum threshold.  This means that in our results, the most
important part of the energy distribution $f(\varepsilon_i)$, which is
peaked around 9, 10 MeV, does not contribute.  Our results are
strongly related to the high energy tails of the $f(\varepsilon_i)$
distributions which we emphasized by using the logarithmic scale in
the insert of the panel (c).

\section{Conclusions}
In this article we have presented the results of calculations for the
electron neutrino, and antineutrino, scattering cross sections off
\car and \oxy nuclei.  The neutrino energy range of our interest goes
from a few tens of MeV up to a few hundreds of MeV. The nuclear
transitions, of both charge-conserving and charge-exchange type, have
been described within the CRPA theory implemented by a model which
considers the FSI effects. The parameters of our nuclear model have
been fixed to reproduce total photo-absorption cross section data,
Fig.  \ref{fig:photf}, and the validity of the model has been tested
against electron scattering data, Fig. \ref{fig:eec12}.

In all the kinematics we have studied, the transverse axial vector
responses give the largest contributions to the cross sections, 
Fig.\ref{fig:resp}.  We have also investigated the angular
distributions of the various cross sections as a function of the
projectile energy, Fig. \ref{fig:ang}.  We found that at low
energies, $\varepsilon_i$ = 50 MeV, all the cross sections are backward
peaked. This behavior is smoothly modified, and already at 200 MeV all
the cross sections are forward peaked.  For energies above 200 MeV
this behavior becomes more accentuated, i.e. the ratio between the
cross section values at forward and backward directions increases.

We have found large differences between neutrinos and electron scattering
cross sections calculated for the same kinematic conditions in the
giant resonance region, Fig. \ref{fig:comp}. The various probes
excite resonances which differ in energy, angular momentum and parity.
This was expected in the charge-exchange processes, since, in this
case, the particle-hole configuration space is different from
the electron scattering reaction. We have found remarkable that
also the charge-conserving neutrino, and antineutrino, cross sections
show different shapes with respect to those of the electron scattering
cross sections. We identified the source of these differences with the
dominance of the transverse axial vector terms in the neutrino
reactions.

Our main results, the total cross sections for both nuclei as a
function of the projectile energy, have been presented in Figs.
\ref{fig:totc12} and \ref{fig:toto16}. For neutrino energies above 40
MeV the effect of the FSI is a reduction of the cross sections of
about 10-20\%, depending upon the nucleus and the reaction considered.
For lower neutrino energies the FSI effects do not have a common trend,
Fig. \ref{fig:ratio}.

The sensitivity of our results to the residual interaction is weak for
neutrino energies above 40 MeV. The various cross sections agree
within a 10\% of accuracy. In this region, we found agreement also
with cross sections calculated bu other authors
\cite{kol99,kol95,kol02,kur90}. This gives us confidence about the
reliability of the nuclear model.

The situation for neutrino energies below 40 MeV is more complicated.
In this case, the cross sections are sensitive only to the excitation
region around the nucleon emission threshold and the giant resonances.
This excitation energy region, is very difficult to describe, since
the CRPA results both depend strongly on the ingredients of the
nuclear structure models, i.e. single particle energies and residual
interactions.  Unfortunately, this problem is not irrelevant for
neutrino physics. We have shown that the energy distributions of the
neutrinos produced by muon decay at rest, see Fig. \ref{fig:michel},
and also that of the neutrinos emitted from a supernova remnant, Fig.
\ref{fig:fluence} are most sensitive to the low energy neutrinos.
Also in the proposed beta-beam facilities \cite{mez03}, the neutrino
energy distributions have their maxima in the region emphasizing the
excitation of nuclear giant resonances \cite{mcl04}.

From the nuclear structure point of view, the situation is quite
intriguing. An improvement of our capability of describing the nuclear
excitation in the continuum is necessary. From the experimental point
of view, (e,e') data in the giant resonance region for the \car and
\oxy nuclei are required to test the validity of our nuclear structure
models. However, being able to describe these data would not be
sufficient to be sure about the description of neutrino-induced
excitations. We have shown that, even in the neutral-current case,
neutrinos calculations are sensitive to nuclear structure details that
are irrelevant in electromagnetic excitations.  Therefore, it is
evident the need of comparisons with hadron excited giant resonance
data \cite{bak97}, especially to test charge-exchange results.

\section*{Acknowledgments}
We thank F.Cavanna and F.Vissani for useful discussions, and
A.M.Lallena and P. Rotelli for their interest in the work and for
their comments about the manuscript. This work has been partially
supported by the MIUR through the PRIN {\sl Fisica del nucleo atomico
  e dei sistemi a molticorpi}.


%
\clearpage
\newpage
%
%
\vskip 3.0 cm
\begin{table}[ht]
\begin{center}
\begin{tabular}{ccccccccc}
\hline
\multicolumn{2}{c}{} & $V$ & $a$ & $R$ & $V_{s}$ & $a_{s}$& $R_{s}$ & 
$R_{c}$ \\ 
\multicolumn{2}{c}{} & [MeV] & [fm] & [fm] & [MeV] & [fm] & 
[fm] & [fm] \\ 
\hline
$^{12}$C & $p$ & $-55.00$ & $0.50$ & $2.95$ & $-4.00$ & $0.50$ & $2.95$ & $%
2.95$ \\ \cline{2-9}
& $n$ & $-55.00$ & $0.50$ & $2.95$ & $-5.50$ & $0.50$ & $2.95$ & $-$ \\ 
\hline
$^{16}$O & $p$ & $-52.50$ & $0.53$ & $3.2$ & $-7.00$ & $0.53$ & $3.20$ & $%
3.20$ \\ \cline{2-9}
& $n$ & $-52.50$ & $0.53$ & $3.2$ & $-6.54$ & $0.53$ & $3.20$ & $-$ \\ 
\hline
\end{tabular}
\end{center}
\bigskip
\caption{\small 
Parameters of Wood-Saxon type potential, see Eqs. 
(\ref{eq:ws1},\ref{eq:ws2},\ref{eq:ws3}),
for \car and \oxy. The label  
$p$ refers to protons and $n$ to neutrons.}
\label{tab:woodpar}
\end{table}
%
%
%
%
\begin{table}[ht]
\begin{center}
\begin{tabular}{clrrrr}
\hline
\multicolumn{2}{c}{ } &
\multicolumn{2}{c}{\car} & 
\multicolumn{2}{c}{\oxy} \\
\hline
\multicolumn{2}{c}{ } & WS   & exp  &  WS   & exp \\ 
\hline
 p  & 1s1/2 & -30.90 &       & -29.90 &  \\
    & 1p3/2 & -15.76 &-15.96 & -16.92 & -18.44 \\
\cline{3-4} 
    & 1p1/2 & -12.95 & -1.94 & -12.74 & -12.11 \\
\cline{5-6}
    & 1d5/2 &   2.0  &       &  -3.76 &  -0.60 \\
    & 2s1/2 &        &       &  -0.89 &  -0.10 \\
\hline
 n  & 1s1/2 & -34.04 &        & -33.98  &  \\
    & 1p3/2 & -18.92 & -18.72 & -20.52  & -21.81 \\
\cline{3-4}
    & 1p1/2 & -15.06 &  -4.96 & -16.63  & -15.65 \\
\cline{5-6}
    & 1d5/2 &  -3.88 &  -1.10 &  -6.84  &  -4.14 \\
    & 2s1/2 &  -1.96 &  -1.86 &  -3.90  &  -3.27 \\
    & 1d3/2 &    1.6 &        &  -3.90  &  -3.27 \\
\hline
\end{tabular}
\end{center}
\bigskip
\caption
{\small 
Single particle energies in MeV for protons (p) and neutrons 
(n) states. 
The label WS indicates the theoretical results. The experimental
energies have been deduced from the level schemes of the neighboring
nuclei \protect\cite{led78}.
The horizontal lines indicate the Fermi surface. 
This is the discrete part of the configuration space used in the CRPA
calculations.  
}
\label{tab:spe}
\end{table}
%
%
\vskip 3.0 cm
\begin{table}[ht]
\begin{center}
\begin{tabular}{l r r r r r r c c}
\hline
        & $F^{ext}$ &  $F^{int}$  &  ${F'}^{\,ext}$  
        &  ${F'}^{\,int}$ 
        &  $G$    &  $G'$     
        & $\mathcal {K}$ ($^{12}$C) & $\mathcal {K}$ ($^{16}$O) \\
\hline
LM1 &  -740.0 & 60.4 & 453.0 & 453.0 & 166.1 & 211.4 & 0.8 & 0.865  \\
LM2 &  -630.0 &  0.0 & 450.0 & 195.0 &   0.0 & 300.0 & 1.0 & 0.865  \\
PP  & -1007.0 &-54.7 & 677.8 & 201.3 &  60.0 & 241.0 & 1.2 & 1.030  \\
\hline
\end{tabular}
\end{center}
\bigskip
\caption
{\small Values of the parameters, in MeV fm$^3$,
   of the three forces used in our calculations, see Eq. (\ref{eq:force}). 
 The $\mathcal K$ factors renormalize the forces. 
}
\label{tab:force}
\end{table}
%
%
%
%
\begin{table}[ht]
\begin{center}
\begin{tabular}{l c c c }
\hline
    & \car$\rightarrow$\car  & \car$\rightarrow^{12}$B &
    \car$\rightarrow^{12}$N \\ 
\hline
LM1 & 13.80 & 11.28 & 17.08 \\
LM2 & 13.89 & 11.36 & 17.19 \\
PP  & 12.94 & 11.23 & 17.03 \\
\hline
exp & 15.11 & 13.37 & 17.34 \\
\hline
\end{tabular}
\end{center}
\bigskip
\caption
{\small Excitation energies, in MeV, of the 1$^+$ states in \car
for charge-conserving excitation and charge-exchange excitations.
Experimental energies are from Ref. \protect\cite{led78}.
}
\label{tab:oneplus}
\end{table}
%
%
%
\vskip 3.0 cm
\begin{table}[ht]
\begin{center}
\begin{tabular}{c c c c c}
\hline
$\varepsilon$ [MeV]&  $A$ [MeV] &  $B$ [MeV]& $C$ [MeV]& $m^* / m $ \\
\hline
      10.0 & 6.0  & 60.0 & 110.0 & 1.0 \\ 
      40.0 & 11.0  & 20.0 & 110.0 & 0.85 \\ 
\hline
\end{tabular}
\end{center}
\bigskip
\caption
{\small Parameters of the function $\gamma ( \varepsilon)$, Eq.
 (\protect\ref{eq:sgamma}), and of the nucleon effective mass.  
 The values of the upper row have been used 
 for $\varepsilon \le$ 10 MeV. Those of the lowest row for $\varepsilon
 \ge$ 40 MeV. For intermediate energies we used values
 obtained by making a linear interpolation between those given in
 the table.
}
\label{tab:gamma}
\end{table}
%
%
%
\begin{table}[ht]
\begin{center}
\begin{tabular}{c c c c c c}
\hline
$J^\Pi$ &  $(e,e')$ &  $(\nu,\nu')$ & 
$(\overline{\nu},\overline{\nu}')$ & $(\nu,e^-)$ & 
$(\overline{\nu},e^+)$ \\
\hline
 $\theta = 30^o$ &  &  &  &  &  \\ 
 1$^-$ & 0.93     & 0.47 & 0.49 & 0.30   & 0.28 \\ 
 1$^+$ & $<$ 0.01 & 0.04 & 0.04 & 0.03   & $<$0.01 \\ 
 2$^-$ & $<$ 0.01 & 0.43 & 0.39 & 0.57   & 0.69 \\ 
 2$^+$ & 0.05     & 0.02 & 0.04 & $<$0.01& $<$0.01 \\ 
 {\sl oth} & 0.02     & 0.04 & 0.04 & 0.10 & 0.03 \\ 
\hline
 $\theta = 150^o$ &  &  &  &  &  \\ 
 1$^-$ & 0.63 & 0.48        & 0.50        & 0.49        & 0.28 \\ 
 1$^+$ & 0.01 & 0.04        & 0.04        & 0.03        & 0.02 \\ 
 2$^-$ & 0.15 & 0.43        & 0.39        & 0.45        & 0.63 \\ 
 2$^+$ & 0.18 & 0.02        & 0.04        & 0.01        & 0.03 \\ 
 {\sl  oth}  & 0.03 & 0.03        & 0.03        & 0.02        & 0.04 \\ 
\hline
\end{tabular}
\end{center}
\bigskip
\caption
{\small Percentage contribution of the various multipoles to the 
  energy integrated cross
  sections of Fig. \protect\ref{fig:comp}. The raws labeled {\sl oth}
  give the contribution of the multipoles other than those labeled in
  the Table up to J=12. 
}
\label{tab:mul}
\end{table}
%
%
\begin{table}[ht]
\begin{center}
\begin{tabular}{l r r }
\hline
    & RPA &  FSI  \\
\hline
LM1 & 5.17 & 4.48  \\
LM2 & 8.12 & 7.15  \\
PP  & 7.70 & 5.77  \\
\hline
    & \multicolumn{2}{c}{14.1$\pm$1.6$\pm$1.9} \protect\cite{kra92} \\
exp & \multicolumn{2}{c}{14.8$\pm$0.7$\pm$1.4} \protect\cite{ath92} \\
    & \multicolumn{2}{c}{14.0$\pm$1.2} \protect\cite{bod94} \\
\hline
\end{tabular}
\end{center}
\bigskip
\caption
{\small Total \car$(\nu,e^-)^{12}N$  flux averaged cross sections, 
  in 10$^{-16}$ fm$^2$ units, for
  electron neutrinos emitted from muon decay at rest.
}
\label{tab:cross}
\end{table}
%
%
%
\begin{table}[ht]
\begin{center}
\begin{tabular}{c c c c}
\hline
    & T [MeV] &  $\alpha$ & $<\varepsilon_i>$ [MeV]  \\
\hline
I    & 3.49 & 0.0 & 11.0  \\
II   & 3.80 & 0.0 & 12.0  \\
III  & 2.46 & 4.0 & 11.0 \\
\hline
\end{tabular}
\end{center}
\bigskip
\caption
{\small Parameters used in the energy distribution $f(\varepsilon_i)$,
  Eq. (\protect\ref{eq:fluence}), and the respective average energies.
}
\label{tab:fluence}
\end{table}
%
\newpage
%
%
%
%
%
%
%
%
%
%
%
%
%
\end{document}